\documentclass[10pt,twocolumn,twoside,]{IEEEtran}
\usepackage{ntheorem}
\usepackage{amsmath}
\usepackage{amsfonts}
\newtheorem{theorem}{Theorem}
\newtheorem{corollary}{Corollary}
\newtheorem{lemma}{Lemma}
\newtheorem{definition}{Definition}
\newtheorem{proposition}{Proposition}

\newtheorem{remark}{Remark}
\usepackage{ntheorem}
\usepackage{amsmath}
\usepackage{amsfonts}
\usepackage{listings}
\usepackage{color}
\usepackage{array}
\usepackage{tabu}
\usepackage{mathrsfs,amssymb}
\usepackage{graphicx}
\graphicspath{ {figure/} }
\usepackage{float}
\usepackage{algorithm}

{}

\usepackage{algpseudocode}
\algdef{SE}[DOWHILE]{Do}{doWhile}{\algorithmicdo}[1]{\algorithmicwhile\ #1}%
\DeclareMathOperator*{\diag}{diag}
\begin{document}
	
	\title{Risk Assessment for Connected Vehicles under Stealthy Attacks on Vehicle-to-Vehicle Networks}
	
	%
	%
	%
	
	\author{Tianci Yang\textsuperscript{1}, Carlos Murguia\textsuperscript{2}, and Chen Lv\textsuperscript{1}
		\thanks{This work was supported by the SUG-NAP Grant (No. M4082268.050) of Nanyang Technological University, Singapore.}
	\thanks{\textsuperscript{1} T. Yang and C. Lv are with the School of Mechanical and Aerospace Engineering, Nanyang Technological University, Singapore. Emails:
		\{tianci.yang, lyuchen\}@ntu.edu.sg}%
	\thanks{\textsuperscript{2} C. Murguia is with the Department of Mechanical Engineering, Eindhoven University of Technology, The Netherlands. Email:
		c.g.murguia@tue.nl}
	}
	%
	%

	\markboth{Journal of \LaTeX\ Class Files,~Vol.~14, No.~8, April~2020}%
	{Shell \MakeLowercase{\textit{et al.}}: Bare Demo of IEEEtran.cls for IEEE Journals}
	%



	\maketitle
	
\begin{abstract}
Cooperative Adaptive Cruise Control (CACC) is an autonomous vehicle-following technology that allows groups of vehicles on the highway to form in tightly-coupled platoons. This is accomplished by exchanging inter-vehicle data through Vehicle-to-Vehicle (V2V) wireless communication networks. CACC increases traffic throughput and safety, and decreases fuel consumption. However, the surge of vehicle connectivity has brought new security challenges as vehicular networks increasingly serve as new access points for adversaries trying to deteriorate the platooning performance or even cause collisions. In this manuscript, we propose a novel attack detection scheme that leverage real-time sensor/network data and physics-based mathematical models of vehicles in the platoon. Nevertheless, even the best detection scheme could lead to conservative detection results because of unavoidable modelling uncertainties, network effects (delays, quantization, communication dropouts), and noise. It is hard (often impossible) for any detector to distinguish between these different perturbation sources and actual attack signals. This enables adversaries to launch a range of attack strategies that can surpass the detection scheme by hiding within the system uncertainty. Here, we provide risk assessment tools (in terms of semidefinite programs) for Connected and Automated Vehicles (CAVs) to quantify the potential effect of attacks that remain hidden from the detector (referred here as \emph{stealthy attacks}). A numerical case-study is presented to illustrate the effectiveness of our methods.
	\end{abstract}
	
	\section{Introduction}
	In the last few decades, highway capacity has become increasingly limited, causing severe traffic congestion problems to our society. An effective way to enhance road capacity is to decrease the intervehicle distance. Cooperative Adaptive Cruise Control (CACC) is a vehicular technology that allows groups of vehicles to form in tightly-coupled platoons by exchanging intervehicle data through Vehicle-to-Vehicle (V2V) wireless communication networks. CACC schemes are able to achieve string stability \cite{ploeg2011design}\cite{swaroop1996string}\cite{ploeg2013lp} (decreasing the effect of disturbances throughout the vehicle string) and increase traffic throughput \cite{van2006impact}. However, the surge of vehicles exposure to the cyber world has brought new security challenges since wireless vehicular networks increasingly serve as new access points for adversaries trying to threaten traffic safety \cite{El-rewini2020}\nocite{laurendeau2006threats}\nocite{miller2015remote}\nocite{Creenberg}\nocite{amoozadeh2015security}\nocite{petit2014potential}\nocite{raya2007securing}-\cite{amoozadeh2015security}. In \cite{Creenberg}, the authors show that they can remotely disable the brakes and commandeer the steering wheel of some Chrysler vehicles. This brought a great financial loss to Chrysler as they had to issue a recall of 1.4 million vehicles. It is of importance for society to realize that cyberattacks to connected vehicles severely threaten human lives since one vehicle hack might lead to fatalities of not only its driver and passengers, but also the pedestrians and drivers and passengers of the other vehicles. Hence, strategic mechanisms are needed for coping with cyberattacks on connected vehicles.
	
Most of existing results in the literature on security of connected vehicles are based on cryptography techniques. For instance, in \cite{raya2007securing}\cite{Chen2019}\cite{mundhenk2017security}\cite{8939382}, different authentication and authorization protocols are provided for preventing in-vehicle or inter-vehicle networks from being attacked. Results that focus on mitigating performance degradation induced by attackers are essential but still quite rare. In \cite{Liu2019,wang2020real}, detection and isolation algorithms for a single vehicle under sensor attacks are provided by exploiting sensor redundancy. The authors of \cite{merco2018replay} provide suitable countermeasures to detect replay attacks for connected vehicles. Mousavinejad et al. provide an algorithm for detecting sensor attacks on connected vehicles using a set-membership filtering technique \cite{mousavinejad2019distributed}. In \cite{ju2020deception}, the problem of attack detection and estimation for connected vehicles under sensor attacks is solved using an unbiased finite impulse response (UFIR) estimator. The problem of attack detection and estimation for vehicle platoons is addressed in \cite{ju2020distributed}-\nocite{9485094}\cite{9502914}.
	
Connected and Automated Vehicles (CAVs) can be equipped with standard fault/attack detectors for identifying faults and attacks \cite{lopes2020active}-\nocite{wang2020anomaly}\nocite{hwang2009survey}\cite{isermann2005fault}. The main idea behind detection schemes is to compute the difference between sensor measurements and estimated outputs provided by a system \textit{estimator} (an algorithm that produce estimates of sensor data based on available information). Alarms are triggered if the difference is larger than a predefined threshold. This standard detection procedure might be useful for detecting system faults or false data injected by adversaries at the communication network. A degradation from CACC to \emph{non-cooperative} Adaptive Cruise Control (ACC) might be an countermeasure when alarms are triggered so that more severe consequences such as vehicle crashes can be prevented \cite{amoozadeh2015security}. That is, if attacks on the communication network are detected, we could simply disconnect the vehicle from that network and switch to a non-cooperative platooning scheme (ACC). This idea works well under the assumption that attacks can be actually detected. However, even the best detection scheme leads to conservative detection results because of unavoidable modelling uncertainties, network effects (delays, quantization, communication dropouts), and noise. It is hard (often impossible) for any detector to distinguish between these different perturbation sources and actual attack signals. This enables adversaries to launch a range of attack strategies that can surpass the detection scheme by hiding within the system uncertainty. In this case, a fundamental question arises: how does a given attack detector constrain what the attacker can inject while remaining undetected? More specifically, is it possible for the attacker to drive the vehicle to a dangerous (unsafe) situation by injecting stealthy attacks on the communication network?
There are various approaches for analyzing the impact of stealthy FDI attacks on general Cyber-Physical Systems (CPSs) \cite{murguia2020security}-\nocite{murguia2017reachable}\nocite{mo2015performance}\nocite{pasqualetti2013attack}\nocite{milovsevic2019estimating}\nocite{kafash2018constraining}\cite{liu2021reachability}. In particular, they characterize the system trajectories (what values can physical variables take) that stealthy attacks can induce in the CPS.

In this manuscript, for platoons of CAVs, we provide mathematical tools, in terms of semidefinite programs, to quantify how stealthy attacks on V2V communication networks can potentially deteriorate the platooning performance, given a class of attack detectors (observer-based robust detectors). In particular, we consider the set of vehicle states (relative positions, velocities, and accelerations) that stealthy attacks can induce (the attacker's reachable set) and use the ``size'' of this set as a security metric for each CAV in the platoon. Since computing this set exactly is not computationally tractable \cite{murguia2020security}, we use ellipsoidal sets to over approximate reachable sets. That is, we compute an outer ellipsoidal approximation of the attacker's reachable set (obtained by solving a set of semidefinite programs) and use its size (in terms of volume) to approximate the security metric.
As a second security metric, we use the minimum distance between the attacker's reachable set and a set of vehicle critical states -- states that if reached something wrong will happen to the vehicle, e.g., crashes. We use the distance between the outer ellipsoidal approximation and the set of critical states to approximate this second security metric. This distance is used for security assessment. It tells us whether there exist attack signals that can drive our vehicle to a collision while remaining undetected to the equipped attack detector. In general, the framework that we provide in this manuscript provide risk assessment tools to quantify how vulnerable a CAV is to stealthy attacks given a CACC scheme and a particular attack detector. The framework can also be used as a guidance for redesigning the controller and monitor so that the impact of stealthy attacks on platooning performance can be reduced and severe consequences such as vehicle crashes can be avoided.

There are a few results in this direction already. In \cite{dadras2018reachable}, the authors consider that several vehicles form a platoon in a non-cooperative manner and there is one malicious vehicle trying to disrupt the platooning dynamics by randomly modifying its own acceleration. Reachability analysis is provided by assuming the modified acceleration is bounded with known bounds. The authors in \cite{sun2020impacts} provide reachability analysis approach for CAVs under FDI attacks and message delay attacks. We remark that the problems considered in \cite{dadras2018reachable} and \cite{sun2020impacts} are fundamentally different from ours since the injected false data is \textit{randomly} constructed without using model knowledge, which can be detected by standard fault detectors and hence those attacks are, by definition, not stealthy. Moreover, no disturbances are considered in \cite{dadras2018reachable} and \cite{sun2020impacts} even though noise and uncertainty in vehicle sensors and V2V communication networks are unavoidable. Also, the analysis approach provided in \cite{dadras2018reachable} only works for the case when the number of vehicles in the platoon is small (because their scheme is centralized). We remark here that our analysis tools are distributed, i.e., no centralized information from all vehicles in the platoon is required. Reachability analysis techniques have been widely used for predicting the behavior of traffic so that safety and efficiency can be enhanced \cite{jiang2020ensuring}-\nocite{wang2021combining}\nocite{shetty2019predicting}\cite{ding2008reachability}.

The main contributions of this manuscript are as follows: 1) we consider vehicle sensors and communication networks that are subject to peak bounded unknown disturbances, and propose two security metrics for quantifying how \textit{stealthy} FDI attacks on V2V communication networks can potentially affect platooning performance; 2) an estimator is designed for each CAV in the platoon, capable of providing Input-to-State Stable (ISS) estimates with respect to disturbances (this estimator can be used for general estimation problems and detecting attacks on V2V communication networks); 3) an estimator-based attack detector is developed for each CAV in the platoon for detecting system faults or general (non stealthy) FDI attacks (as stealthy attacks are by definition undetectable).

The paper is organized as follows. In Section \ref{pre}, some preliminary results needed for the subsequent sections are presented. In Section \ref{sysd}, the considered vehicle platoon system is described. In Section \ref{estimation}, we provide tools for designing state estimators that provide ISS estimates for each CAV in the platoon, and an estimator-based attack detection strategy is developed. A method for computing outer ellipsoidal approximations of attacker's reachable sets is provided in Section \ref{reachable}. In Section \ref{cri}, we provide an approach for obtaining the distance between the outer ellipsoidal approximation and the set of critical states. A numerical example is given in Section \ref{simulation} to demonstrate the performance of our tools. Finally, in Section \ref{conclusion}, concluding remarks are given.

	\section{Preliminaries}\label{pre}
	\subsection{Notation}
	We denote the set of real numbers by $\mathbb{R}$, the set of natural numbers by $\mathbb{N}$, and $\mathbb{R}^{n\times m}$ the set of $n\times m$ matrices for any $m,n \in \mathbb{N}$. For any vector $v\in\mathbb{R}^{n}$,  we denote  $|v|=\sqrt{v^{\top} v}$. For a signal $\left\lbrace v(t)\right\rbrace _{t=0}^{\infty}$,  $||v||_{\infty} := \sup_{t\geq 0}|v(t)|$, $v(t) \in \mathbb{R}^{n}$. We say that a signal $\left\lbrace v(t)\right\rbrace$ belongs to $l_{\infty}$, $\left\lbrace v(t)\right\rbrace \in l_{\infty}$, if $||v||_{\infty} <\infty$. $||v(t)||_{\mathcal{L}_{p}}$ is the p-norm of signal $v(t)$.

	\subsection{Definitions and Mathematical Preliminaries}

	\begin{definition}[Reachable Set]\emph{\cite{murguia2020security}.}
		Consider the perturbed LTI discrete-time system:\label{def1}
		\begin{equation}\label{s1}
			\xi(k+1)=\mathcal{A}\xi(k)+\sum_{i=1}^{N}\mathcal{B}_{i}w_{i}(k),
		\end{equation}
		with time-step $k \in \mathbb{N}$, state $\xi \in \mathbb{R}^{n_{\xi}}$, perturbation $w_{i}\in\mathbb{R}^{p_{i}}$ satisfying $w_{i}^{\top}W_{i}w_{i}\leq 1$, for some positive definite matrix $W_{i}\in\mathbb{R}^{p_{i}\times p_{i}}$, $i=1,\ldots,N$, and matrices $\mathcal{A}\in\mathbb{R}^{n_{\xi}\times n_{\xi}}$ and $\mathcal{B}_{i}\in\mathbb{R}^{n_{\xi}\times p_{i}}$, $N, n_\xi, p_i\in\mathbb{N}$. The reachable set $\mathcal{R}_{k}^{\xi}$, at time-step $k \in \mathbb{N}$, from the initial condition $\xi(1)$ is the set of states reachable in $k$ steps by system \eqref{s1} through all possible perturbations satisfying $w_{i}^{\top}W_{i}w_{i}\leq 1$, i.e.,
		\begin{equation}
			\mathcal{R}^{\xi}_{k}:=\left\lbrace \xi\in\mathbb{R}^{n_{\xi}}\vline\hspace{1mm}\xi\hspace{1mm}\text{satisfy \eqref{s1}}, \text{and}\hspace{1mm} w_{i}^{\top}W_{i}w_{i}\leq 1\right\rbrace.
		\end{equation}
	\end{definition}

	\begin{lemma}[Ellipsoidal Approximation]\emph{\cite{murguia2020security}}. Consider the perturbed LTI system \eqref{s1} and the reachable set $\mathcal{R}_{k}^{\xi}$ introduced in Definition \ref{def1}. For a given $a\in(0,1)$, if there exist constants $a_{1}=a_{1}^{*}$, $\ldots$, $a_{N}=a_{N}^{*}$ and matrix $P=P^{*}$ solution of the convex optimization:
		\begin{equation}
			\left\{\begin{split}
				&\min_{P,a_{1},\ldots, a_{N}}-\log\det[P],\\
				&s.t. \hspace{1mm}a_{1},\ldots, a_{N}\in(0,1), a_{1}+\ldots+ a_{N}\geq a,\\
				&P>0, \begin{bmatrix}
					aP&\mathcal{A}^{\top}P&0\\
					P\mathcal{A}&P&P\mathcal{B}\\
					0&\mathcal{B}^{\top}P&W_{a}
				\end{bmatrix}\geq 0,
			\end{split}\right.
		\end{equation}
		with $W_{a}:=\diag\begin{bmatrix}
			(1-a_{1})W_{1},\ldots,(1-a_{N})W_{N}
		\end{bmatrix}\in\mathbb{R}^{\bar{p}\times\bar{p}}$, $\mathcal{B}:=(\mathcal{B}_{1},\ldots,\mathcal{B}_{N})\in\mathbb{R}^{n_{\xi}\times \bar{p}}$, and $\bar{p}=\sum_{i=1}^{N}p_{i}$; then for all $k \in \mathbb{N}$, $\mathcal{R}_{k}^{\xi}\subseteq\mathcal{E}_{k}^{\xi}:=\left\lbrace \xi^{\top}P^{\xi}\xi\leq\alpha_{k}^{\xi}\right\rbrace $, with $P^{\xi}:=P^{*}$ and $\alpha_{k}^{\xi}:=a^{k-1}\xi(1)^{\top}P^{*}\xi(1)+((N-a)(1-a^{k-1}))/(1-a)$ and the ellipsoid $\mathcal{E}_{k}^{\xi}$ has the minimum volume among all outer ellipsoidal approximations of $\mathcal{R}_{k}^{\xi}$. \label{lemma1a}
	\end{lemma}
	\begin{lemma}[Ellipsoid Projection]\emph{\cite{murguia2020security}}.\label{lemma1}
		Consider the following ellipsoid:
		\begin{equation}
			\begin{split}
				\mathcal{E}:=\left\lbrace x\in\mathbb{R}^{n}, y\in\mathbb{R}^{m}\bigg|\begin{bmatrix}
					x\\y
				\end{bmatrix}^{\top} \underbrace{\begin{bmatrix}
					Q_{1}&Q_{2}\\Q_{2}^{\top}&Q_{3}
				\end{bmatrix}}_{Q}\begin{bmatrix}
					x\\y
				\end{bmatrix}=\alpha \right\rbrace ,
			\end{split}
		\end{equation}
		for some positive definite matrix $Q\in\mathbb{R}^{(n+m)\times(n+m)}$ and constant $\alpha\in\mathbb{R}_{>0}$. The projection $\mathcal{E}'$ of $\mathcal{E}$ onto the $x$-hyperplane is given by the ellipsoid:
		\begin{equation}
			\mathcal{E}':=\left\lbrace x\in\mathbb{R}^{n}\vline  \hspace{1mm}x^{\top}\begin{bmatrix}
				Q_{1}-Q_{2}Q_{3}^{-1}Q_{2}^{\top}
			\end{bmatrix}x=\alpha\right\rbrace .
		\end{equation}
	\end{lemma}

	\begin{figure}[t]\centering
		\includegraphics[width=0.5\textwidth]{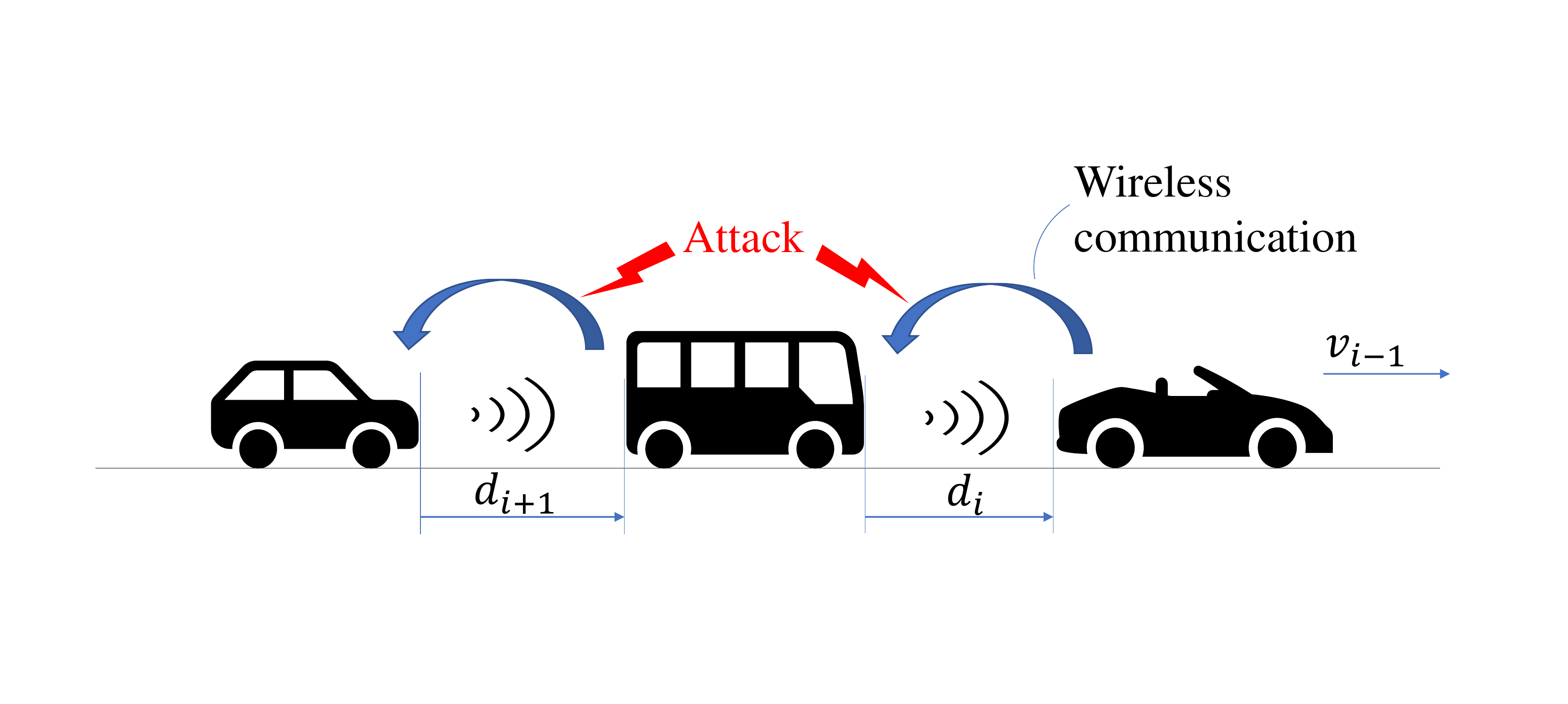}
		\caption{Vehicle platoon under communication network attacks.}
		\centering
		\label{fig:1}
	\end{figure}

	\section{System Description}\label{sysd}
	Consider a platoon of $m$ vehicles as depicted in Figure \ref{fig:1}. Denote the distance between vehicle $i$ and its preceding vehicle $i-1$ as $d_{i}$, and its velocity as $v_{i}$. The objective of each vehicle is to keep a desired distance, $d_{r,i}$, with respect to its preceding vehicle:
	\begin{equation}\label{distance}
		d_{r,i}(t)={s_{i}}+hv_{i}(t), i\in S_{m},
	\end{equation}
	where $h$ is the time-headway constant and {$s_{i}$} is a desired standstill distance. The set $S_{m}=\left\lbrace i\in\mathbb{N}|1\leq i\leq m\right\rbrace $ denotes the set of all vehicles in a platoon of length $m\in\mathbb{N}$. The spacing policy adopted here improves string stability -- attenuation of the effect of disturbances throughout the vehicle string \cite{rajamani2002semi}-\cite{Ploeg2014}. The spacing error $e_{i}(t)$ is defined as
	\begin{equation}\label{e}
		\begin{split}
			e_{i}(t)=&d_{i}(t)-d_{r,i}(t),\\
			=&(q_{i-1}(t)-q_{i}(t)-L_{i})-({s_{i}}+hv_{i}(t)),
		\end{split}
	\end{equation}
	where $q_{i}$ denotes the rear-bumper position of vehicle $i$ and $L_{i}$ denotes its length. We consider the following vehicle model adopted in \cite{Ploeg2014}:
	\begin{equation}\label{vehicle}
		\begin{split}
			\begin{bmatrix}
				\dot{d}_{i}\\
				\dot{v}_{i}\\
				\dot{a}_{i}
			\end{bmatrix}=\begin{bmatrix}
				v_{i-1}-v_{i}\\
				a_{i}\\
				-\frac{1}{\tau}a_{i}+\frac{1}{\tau}u_{i}
			\end{bmatrix},\hspace{2mm}i\in S_{m},
		\end{split}
	\end{equation}
where $\tau$ denotes the driveline dynamics constant, $a_{i}$ is the acceleration of vehicle $i$, and $u_{i}$ is its control input. Note that all vehicles have the same $\tau$ (i.e., we assume the vehicle string is homogeneous as in \cite{Ploeg2014}). We adopt the CACC dynamic controller for $u_i$ introduced in \cite{Ploeg2014}, which fulfills the vehicle-following objective and enforces string stability:
	\begin{equation}\label{control_a}
		\begin{split}
			h\dot{u}_{i}=-u_{i}+\epsilon_{i},
		\end{split}
	\end{equation}
	with
	\begin{equation}\label{control}
		\epsilon_{i} := K\begin{bmatrix}
			e_{i}\\
			\dot{e}_{i}
		\end{bmatrix}+u_{i-1}+\delta_{i}+\omega_{ui},\hspace{1mm}i\in S_{m},
	\end{equation}
	and $K=\begin{bmatrix}
		k_{p}&k_{d}
	\end{bmatrix} \in \mathbb{R}^{1 \times 2}$. The feedforward term $u_{i-1}$ is transmitted from vehicle $i-1$ to vehicle $i$ through a wireless communication channel, which is perturbed by channel noise, network effects, and cyberattacks. Signal $\omega_{ui}\in\mathbb{R}$ encompasses the noise in the communication channel and network-induced imperfections (e.g., quantization and packet dropouts). Signal $\delta_{i}\in\mathbb{R}$ represents the injected attack signal, i.e., $\delta_{i}(t)\neq 0$ for some $t\geq 0$ if the communication network is compromised; otherwise, $\delta_{i}(t)=0$ for all $t\geq 0$. Using \eqref{distance}-\eqref{control}, the following closed-loop platoon model is obtained:
	\begin{equation}\label{ee1}
		\begin{split}
			\begin{bmatrix}
				\dot{e}_{i}\\
				\dot{v}_{i}\\
				\dot{a}_{i}\\
				\dot{u}_{i}
			\end{bmatrix}=&\begin{bmatrix}
				0&-1&-h&0\\
				0&0&1&0\\
				0&0&-\frac{1}{\tau}&\frac{1}{\tau}\\
				\frac{k_{p}}{h}&-\frac{k_{d}}{h}&-k_{d}&-\frac{1}{h}
			\end{bmatrix}\begin{bmatrix}
				{e}_{i}\\
				{v}_{i}\\
				{a}_{i}\\
				{u}_{i}
			\end{bmatrix}\\
			&+\begin{bmatrix}
				0&1&0\\
				0&0&0\\
				0&0&0\\
				\frac{k_{p}}{h}&\frac{k_{d}}{h}&\frac{1}{h}
			\end{bmatrix}\begin{bmatrix}
				\omega_{di}\\
				v_{i-1}+\omega_{vi}\\
				u_{i-1}+\delta_{i}+\omega_{ui}
			\end{bmatrix}, \hspace{3mm} i\in S_{m},
		\end{split}
	\end{equation}
	where $\omega_{di}$, $\omega_{vi},\omega_{ui}\in\mathbb{R}$ denote bounded perturbations/noise in radars, velocity sensors, and communication channels, respectively. Note that in order to implement controller \eqref{control_a}-\eqref{control}, besides data transmitted through V2V networks, $u_{i-1}$, we need realizations of $e_i$ in \eqref{e}, its time derivative $\dot{e}_i = v_{i-1} - v_{i} - ha_i$, and $u_i$. It follows that, in \cite{Ploeg2014}, it is implicitly assumed that $y_i := [e_i,v_i,a_i,u_i,v_{i-1} - v_{i}]^\top$ are available for feedback. Here, we use the same output $y_i$ to drive our algorithms but assume unavoidable noise affecting all these signals.\\ 
	
The first vehicle in the platoon (which does not have preceding vehicle in front) follows a virtual reference vehicle $(i=0)$, so that the same controller used for the other vehicles can be used to drive the lead vehicle. We formulate the virtual reference vehicle as follows:
	\begin{equation}\label{ee2}
		\begin{split}
			\begin{bmatrix}
				\dot{e}_{0}\\
				\dot{v}_{0}\\
				\dot{a}_{0}\\
				\dot{u}_{0}
			\end{bmatrix}=\begin{bmatrix}
				0&0&0&0\\0&0&1&0\\0&0&-\frac{1}{\tau}&\frac{1}{\tau}\\0&0&0&-\frac{1}{h}
			\end{bmatrix}\begin{bmatrix}
				e_{0}\\
				v_{0}\\
				a_{0}\\
				u_{0}
			\end{bmatrix}+\begin{bmatrix}
				0\\0\\0\\\frac{1}{h}
			\end{bmatrix}\epsilon_{0},
		\end{split}
	\end{equation}
	where $\epsilon_{0}$ denotes the external platoon input representing the human driver in the lead vehicle.

For implementing our analysis methods, we let $\tilde{\omega}_{i}=\begin{bmatrix}
		\omega_{di},v_{i-1}+\omega_{vi},u_{i-1}+\omega_{ui}
	\end{bmatrix}^{\top}$, $x_{i}=\begin{bmatrix}
		e_{i}&
		v_{i}&
		a_{i}&
		u_{i}
	\end{bmatrix}^{\top}$, and formulate system \eqref{ee1} in the following compact form:
	\begin{equation}\label{ss}
	\begin{split}
			\dot{x}_{i}=&A_{c}x_{i}+B_{c}\tilde{\omega}_{i}+\Gamma_{c}\delta_{i},
		\end{split}
	\end{equation}
	with
	\begin{equation}\label{matrix11}
		\begin{split}
			A_{c}:=&\begin{bmatrix}
				0&-1&-h&0\\
				0&0&1&0\\
				0&0&-\frac{1}{\tau}&\frac{1}{\tau}\\
				\frac{k_{p}}{h}&-\frac{k_{d}}{h}&-k_{d}&-\frac{1}{h}
			\end{bmatrix},\\
			B_{c}:=&\begin{bmatrix}
				0&1&0\\
				0&0&0\\
				0&0&0\\
				\frac{k_{p}}{h}&\frac{k_{d}}{h}&\frac{1}{h}
			\end{bmatrix},\Gamma_{c}:=\begin{bmatrix}
			0\\0\\0\\\frac{1}{h}
		\end{bmatrix}.\\
		\end{split}
	\end{equation}
We exactly discretize \eqref{ss} at the sampling time instants, $t = T_{s}k$, $k \in \mathbb{N}$, assuming a zero-order hold to implement control actions and model discrete-time uncertainties (see \cite{Astrom} for details), and obtain the equivalent discrete-time systems:
\begin{equation}\label{ssd}
		\begin{split}
			x_{i}(k+1)=&Ax_{i}(k)+B\tilde{\omega}_{i}(k)+\Gamma\delta_i(k),
		\end{split}
	\end{equation}
with $x_{i}(k) := x_{i}(T_{s}k)$, $\tilde{\omega}_{i}(k) := \tilde{\omega}_{i}(T_{s}k)$, $\delta_i(k) := \delta_i(T_{s}k)$, and matrices
	\begin{equation}\label{matrix12}
\left\{
		\begin{split}
			A=&e^{A_{c}T_{s}}, \hspace{1mm} B = \int_{0}^{T_{s}}e^{A_{c}(T_{s}-s)}B_{c}ds,\\
			 \hspace{1mm} \Gamma=&  \int_{0}^{T_{s}}e^{A_{c}(T_{s}-s)}\Gamma_{c}ds.
		\end{split}
\right.
	\end{equation}

In this manuscript, we address and solve the following problems:\\
	\textbf{Problem 1:} \emph{Design an estimator for each CAV in the platoon that provides Input-to-State Stable (ISS) \emph{\cite{sontag2008input}} state estimates with respect to perturbations $\tilde{\omega}_{i}$.}\\
	\textbf{Problem 2:} \emph{Develop a distributed estimator-based attack detection strategy for each CAV in the platoon.}\\
     \textbf{Problem 3:} \emph{Provide tools for quantifying vulnerabilities of CAVs to stealthy attacks given the CACC controller introduced above and the developed attack detection strategy.}

\section{Estimator and Monitor Design}\label{estimation}
We assume an estimator is used by each vehicle to compute state estimates from noisy sensor measurements. Moreover, each CAV in the platoon is equipped with a fault detector to monitor the presence of faults and attacks.

	\subsection{Estimator Design}
We first design an estimator to reconstruct the state $x_i$ of system \eqref{ssd} from onboard noisy sensors $y_i$, local controllers $u_i$, and received data, $u_{i-1}+\delta_{i}+\omega_{ui}$, coming from the preceding vehicle through V2V networks. In particular, the effect of perturbations $\tilde{\omega}_{i}$ on estimation errors is attenuated by 1) designing the estimator such that its error dynamics is Input-to-State Stable \cite{sontag2008input} with respect to perturbations; and 2) minimizing the corresponding ISS gain, see \cite{jiang2001input} for details.
The ultimate goal of the estimator is to be used to detect (or constrain) attacks on V2V networks. To accomplish this, we leverage redundancy between data provided by onboard sensors and data coming from V2V networks. Note that if there is an attack $\delta_i$ injected in the communication between vehicles, \emph{the effect of that attack is not present in the preceding vehicle dynamics}. The controller driving the preceding vehicle is the attack-free $u_{i-1}$ and the data the follower receives is $u_{i-1}+\delta_{i}+\omega_{ui}$ (so there is a mismatch between these signals). Moreover, the onboard sensor measuring relative velocity, $\triangle v_{i}:=v_{i-1}-v_{i}$, carries information about $u_{i-1}$ -- because $\triangle \dot{v}_{i} = a_{i-1} - a_i = \frac{1}{\tau}(u_{i-1}-a_{i-1}) - a_i$. Therefore, by matching what is measured onboard, $\triangle v_{i}$, and what is received through the network, $u_{i-1}+\delta_{i}+\omega_{ui}$, we can spot the presence of attacks. The fundamental question is how we match these two data sources properly to achieve detection given the high level of uncertainty in the system. In this manuscript, to reflect the effect of the attack signal $\delta_{i}$ in the dynamics of the estimator, we consider the extended state $x_{ei}:=\begin{bmatrix}
		e_{i}&
		v_{i}&
		a_{i}&
		u_{i}&
		\triangle v_{i}&
		a_{i-1}
	\end{bmatrix}^{\top}$. Due to the expressions in \eqref{ee1}-\eqref{ee2}, the dynamics of the extended   state is given as follows
\begin{equation}\label{xe}
	\begin{split}
			\dot{x}_{ei}=&\begin{bmatrix}
			0&0&-h&0&1&0\\
			0&0&1&0&0&0\\
			0&0&-\frac{1}{\tau}&\frac{1}{\tau}&0&0\\
			\frac{k_{p}}{h}&0&-k_{d}&-\frac{1}{h}&\frac{k_{d}}{h}&0\\
			0&0&-1&0&0&1\\
			0&0&0&0&0&-\frac{1}{\tau}
		\end{bmatrix}x_{ei}\\
		&+\begin{bmatrix}
			0\\0\\0\\0\\0\\\frac{1}{\tau}
		\end{bmatrix}u_{i-1}+\begin{bmatrix}
			0\\0\\0\\\frac{1}{h}\\0\\0
		\end{bmatrix}(u_{i-1}+\delta_{i}+\omega_{ui}),\\
	=:& A_{ce}x_{ei}+B_{ce1}u_{i-1}+B_{ce2}(u_{i-1}+\delta_{i}+\omega_{ui}),\\
	y_{ei}=&\begin{bmatrix}
		1&0&0&0&0&0\\
		0&1&0&0&0&0\\
		0&0&1&0&0&0\\
		0&0&0&1&0&0\\
		0&0&0&0&1&0
	\end{bmatrix}x_{ei}+\omega_{ei},\\
=:& C_{e}x_{ei}+\omega_{ei}.
	\end{split}
\end{equation}
with $\omega_{ei}$ denoting the vector of noise. Dynamics \eqref{xe} is exactly discretized at the sampling time instants, $t = T_{s}k$, $k \in \mathbb{N}$, assuming again a zero-order hold to implement control actions and model discrete-time uncertainties, and obtain the equivalent discrete-time dynamics:
\begin{equation}\label{xed}
	\left\{\begin{split}
		x_{ei}(k+1)=&A_{e}x_{ei}(k)+B_{e1}u_{i-1}(k)\\ &+B_{e2}(u_{i-1}(k)+\delta_{i}(k)+\omega_{ui}(k)),\\
		y_{ei}(k)=&C_{e}x_{ei}(k)+\omega_{ei}(k),
	\end{split}\right.
\end{equation}
with $x_{ei}(k) := x_{ei}(T_{s}k)$, ${u}_{i}(k) := {u}_{i}(T_{s}k)$, ${\omega}_{ui}(k) := {\omega}_{ui}(T_{s}k)$, ${\omega}_{ei}(k) := {\omega}_{ei}(T_{s}k)$, $\delta_i(k) = \delta_i(T_{s}k)$, and matrices
\begin{equation}
\left\{
	\begin{split}
		A_{e}=&e^{A_{ce}T_{s}}, \hspace{1mm} B_{e1}= \int_{0}^{T_{s}}e^{A_{ce}(T_{s}-s)}B_{ce1}ds,\\
		B_{e2}=&  \int_{0}^{T_{s}}e^{A_{ce}(T_{s}-s)}B_{ce2}ds.
	\end{split}
\right.
\end{equation}
Consider the following estimator:
	\begin{equation}\label{sse}
		\begin{split}
		\hat{x}_{ei}(k+1) &=A_{e}\hat{x}_{ei}(k) + B_{e}\big(u_{i-1}(k)+\delta_{i}(k)+\omega_{ui}(k)\big)\\
		&+L\big(y_{e}(k+1)-C_{e}A_{e}\hat{x}_{ei}(k) \big)\\ &-LC_{e}B_{e}\big(u_{i-1}(k)+\delta_{i}(k)+\omega_{ui}(k)\big),
		\end{split}
	\end{equation}
where $\hat{x}_{ei}(k)\in\mathbb{R}^{6}$ is the state estimate of $x_{ei}$, $L\in\mathbb{R}^{6 \times 2}$ is the estimator gain, $B_{ce}:=\begin{bmatrix}
	0&0&0&\frac{1}{h}&0&\frac{1}{\tau}
\end{bmatrix}^{\top}$, and $B_{e}:=\int_{0}^{T_{s}}e^{A_{ce}(T_{s}-s)}B_{ce}ds$. Note that this estimator is implementable as what the $i$-th vehicle receives from the communication network is $u_{i-1}+\delta_{i}+\omega_{ui}$. We only need this latter quantity and $y_{ei}$ to compute the estimate $\hat{x}_{ei}$ recursively using \eqref{sse}. Define the estimation error $e_{ei}:=x_{ei}-\hat{x}_{ei}$. Given the system dynamics \eqref{xe} and the estimator \eqref{sse}, the estimation error dynamics is described by the following difference equation:
	\begin{equation}\label{sseee}
	\begin{split}
		e_{ei}(k+1)=&\bar{A}e_{ei}(k)-\bar{L}B_{e1}\big(\delta_{i}(k)+\omega_{ui}(k)\big)\\ &-L\omega_{ei}(k+1),
	\end{split}
\end{equation}
	with
\begin{equation*}
	\bar{A} :=(I-LC_{e})A_{e} \text{ and } \bar{L} :=I-LC_{e}.
\end{equation*}
	Since $C_{e}B_{e1}=\mathbf{0}$, we have $\bar{L}B_{e1}=B_{e1}$ and
	\begin{equation}\label{sseee}
		\begin{split}
			e_{ei}(k+1)=&\bar{A}e_{ei}(k)-B_{e1}\big(\delta_{i}(k)+\omega_{ui}(k)\big)-L\omega_{ei}(k+1).
		\end{split}
\end{equation}
\begin{remark}
	Because the estimation error dynamics \eqref{sseee} depends explicitly on the attack signal $\delta_{i}$, the estimator can be used to detect attacks on V2V communication networks. However, if the estimator is designed for the original system \eqref{ssd} without adding the dynamics of $\triangle v_{i}$ and $a_{i-1}$, it can be verified that the corresponding estimation error dynamics would be completely independent of $\delta_{i}$. This indicates that such an (standard) estimator designed for \eqref{ssd} cannot be used for detecting faults or attacks on V2V communication networks. We need to extend the dynamics with $\triangle v_{i}$ and $a_{i-1}$ to capture the effect of these attacks.
\end{remark}
	Then, in the attack-free case, i.e., $\delta_{i}=0$, we have
	\begin{equation}\label{ssee1}
	\begin{split}
		e_{ei}(k+1)=&\bar{A}e_{ei}(k)-B_{e1}\omega_{ui}(k)-L\omega_{ei}(k+1).
	\end{split}
\end{equation}

In the following lemma, we give a result (in terms of semidefinite programs) to design the estimator gain $L$. The obtained $L$ minimizes the effect of perturbations on the estimation error in the attack-free case.

	\begin{lemma}\label{lemma3}
		Consider the attack-free estimation error dynamics \eqref{ssee1}. For given $\alpha\in(0,1)$, if there exist positive constants $\mu_{1}=\mu_{1}^{*}$ and $\mu_{2}=\mu_{2}^{*}$, and matrices $P=P^{*}$ and $Y=Y^{*}$ solution of the semidefinite program:
			\begin{subequations}\label{lmi3}
	\begin{align}
			&\min_{P,Y}\hspace{2mm}\mu_{1}+\mu_{2},\label{eq:cost}\\
			&s.t. \quad\mu_{1}, \mu_{2}>0, P>\mathbf{0},\\
			&\begin{bmatrix}
				-P&*&*&*\\
				A_{e}^{\top}(P-C_{e}^{\top}Y^{\top})&(\alpha-1) P&*&*\\
				B_{e1}^{\top}P&\mathbf{0}&-\alpha\mu_{1} I&*\\
				Y^{\top}&\mathbf{0}&\mathbf{0}&-\alpha\mu_{1} I
			\end{bmatrix}\leq 0,\label{eq:l1}\\
			&\begin{bmatrix}
				P&I\\
				I&\mu_{2}I
			\end{bmatrix}\geq 0,\label{eq:l2}
		\end{align}
	\end{subequations}
		then, for $L=P^{-1}Y$, the estimation error satisfies:
		\begin{equation}\label{iss}
			\begin{split}
				|e_{ei}(k)|\leq c\lambda^{k}|e_{ei}(0)|+\gamma\big(||\omega_{ui}||_{\infty}+||\omega_{ei}||_{\infty}\big),
			\end{split}
		\end{equation}
		 for some $c>0$, $\lambda\in(0,1)$, $\gamma=\sqrt{\mu_1^{*}\mu_2^{*}}$, and all $k \in \mathbb{N}$.
	\end{lemma}
	\textit{Proof: See Appendix \ref{lm3}.}
\begin{remark}
	By minimizing $\mu_1 + \mu_2$ in \eqref{lmi3}, we are minimizing an upper bound on $\gamma$ in \eqref{iss} by properly selecting the estimator gain $L$. This effectively reduces the effect of perturbations on the estimation error (see \eqref{iss}). Hence, our estimator provides a robust state estimate for CAVs.
\end{remark}

	\subsection{System Monitor Design}
Next, we propose an estimator-based anomaly detector for each CAV in the platoon. This detector quantifies the difference between measured outputs and estimated outputs coming from the estimator \eqref{sse} (the so-called monitoring residuals). If residuals are larger than expected, alarms are triggered and faults/attacks on the vehicle are detected.

Define the system residual as
	\begin{equation}\label{r1}
		\begin{split}
			& r_{i}(k+1) := y_{ei}(k+1)\\ & - C_{e} \big( A_{e}\hat{x}_{ei}(k) + B_{e}(u_{i-1}(k)+\delta_{i}(k)+\omega_{ui}(k)) \big)\\
			&=C_{e}A_{e}e_{ei}(k)+\omega_{ei}(k+1)-C_{e}B_{e1}\delta_{i}(k)-C_{e}B_{e1}\omega_{ui},
		\end{split}
	\end{equation}
where, by construction, $C_{e}B_{e1}=\mathbf{0}$. Then, $r_{i}(k+1)$ evolves according to the difference equation
	\begin{equation}\label{er}
		\left\{\begin{split}	e_{ei}(k+1)=&\bar{A}e_{ei}(k)-B_{e1}\big(\delta_{i}(k)+\omega_{ui}(k)\big)-L\omega_{ei}(k+1),\\
			r_{i}(k+1)=&C_{e}A_{e}e_{ei}(k)+\omega_{ei}(k+1).
		\end{split}
		\right.
	\end{equation}

To detect faults/attacks, we use a quadratic form of the residual, $z_{i}:=r_{i}^{\top}\Pi r_{i}$, for some positive definite matrix $\Pi\in\mathbb{R}^{5\times 5}$. Consider the following monitor.\\
	\begin{algorithm}[h!]
		\textbf{System Monitor:}\\
		\begin{equation}\label{monitor}
			\text{If}\hspace{1mm}z_{i}(k)=r_{i}(k)^{\top}\Pi r_{i}(k)>1, \hspace{1mm}\tilde{k}=k.
		\end{equation}
		\textbf{Design parameter:} positive semidefinite matrix $\Pi\in\mathbb{R}^{2\times 2}$.\\
		\textbf{Output:} alarm time(s) $\tilde{k}$.
	\end{algorithm}

\begin{figure}[t]\centering
	\includegraphics[width=0.5\textwidth]{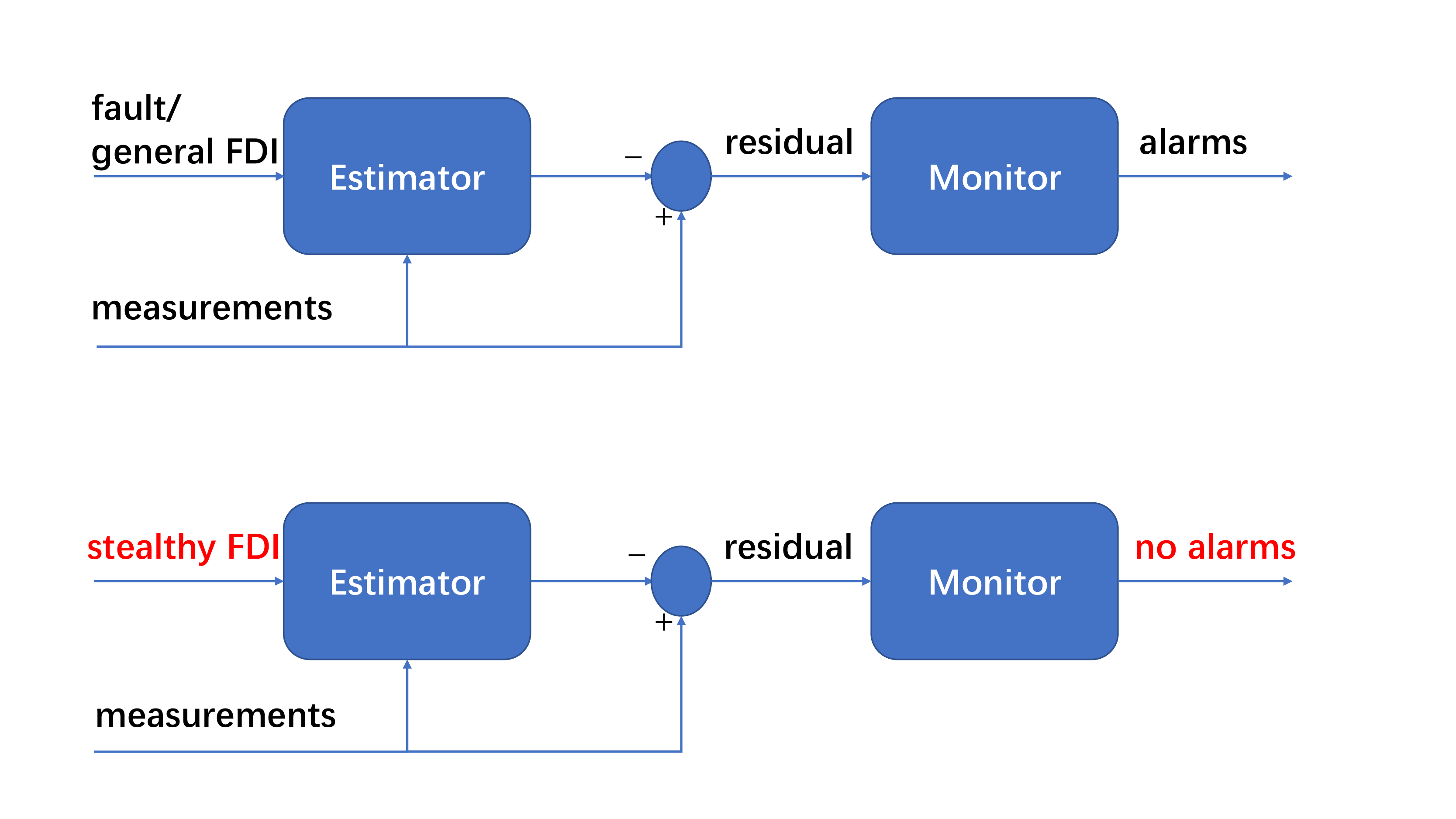}
	\caption{A standard fault/attack detector is useful for identifying a large class of anomalous behaviors including faults or general FDI attacks; but it cannot work in the presence of stealthy FDI attacks.}
	\centering
	\label{fig:x}
\end{figure}

The monitor must be designed such that alarms are triggered if $z_{i}(k)>1$. We select $\Pi$ such that, the ellipsoid $r_{i}(k)^{\top}\Pi r_{i}(k)\leq 1$ contains all the possible trajectories that $\omega_{ui}(k)$ and $\omega_{ei}(k)$ can induce in the residual dynamics \eqref{er} \emph{in steady state} (i.e., after transients due to initial conditions have settle down) given the perturbations bounds $\omega_{ui}(k)^{\top}\omega_{ui}(k)\leq\bar{\omega}_{2}$ and $\omega_{ei}^{\top}(k)\omega_{ei}(k)\leq\bar{\omega}_{3}$, and $\delta_{i}(k)=0$. Note that, the tighter the monitoring ellipsoid, the less opportunity the adversary has to disrupt the system dynamics. Here, we use Lemma \ref{lemma3} and the $S$-procedure \cite{boyd1994linear} to obtain an optimal matrix $\Pi$ (in terms of tightness).
\begin{proposition}
Let the conditions in Lemma \ref{lemma3} be satisfied and consider the associated estimation error bound \eqref{iss} and the residual dynamics \eqref{er}. If there exist $\lambda_{1}$, $\lambda_{2}\in\mathbb{R}_{>0}$ and $\Pi$ solution to the following optimization problem:
	\begin{equation}\label{lmipi}
		\left\{\begin{split}
			&\min_{\Pi,\lambda_{1},\lambda_{2}}-\log\det[\Pi],\\
			&s.t. \hspace{1mm} \Pi>0,\lambda_{1} > 0, \lambda_{2} > 0, \hspace{1mm} \text{and}\\
			&\begin{bmatrix}
				f_{1}&*&*\\
				\Pi C_{e}A_{e}&f_{2}&*\\
				0&0&f_{3}
			\end{bmatrix}	\geq 0,		\\
			&f_{1}=\lambda_{1}I-A_{e}^{\top}C_{e}^{\top}\Pi C_{e}A_{e},\\
			&f_{2}=\lambda_{2}I-\Pi,\\
			&f_{3}=1-\lambda_{1}\gamma^{2}(\bar{\omega}_{2}+\bar{\omega}_{3})-\lambda_{2}\bar{\omega}_{3};
		\end{split}\right.
	\end{equation}
	then, in steady state, the monitor ellipsoid $r_{i}(k)^{\top}\Pi r_{i}(k) = 1$ contains all the trajectories generated by the residual dynamics \eqref{er} for $\delta_{i}(k)=0$ and all $\omega_{ui}(k)$ and $\omega_{ei}(k)$ satisfying $\omega_{ui}(k)^{\top}\omega_{ui}(k)\leq\bar{\omega}_{2}$ and $\omega_{ei}^{\top}(k)\omega_{ei}(k)\leq\bar{\omega}_{3}$.
\end{proposition}
\textit{Proof: See Appendix \ref{p1}.}
\begin{remark}
	By solving \eqref{lmipi}, the matrix $\Pi$ is chosen such that the ellipsoid $r_{i}(k)^{\top}\Pi r_{i}(k)\leq 1$ contains all the possible trajectories that $\omega_{ui}(k)$, $\omega_{ei}(k)$ can induce in the residual given in \eqref{er} with $\omega_{ui}(k)^{\top}\omega_{ui}(k)\leq\bar{\omega}_{2}$, $\omega_{ei}^{\top}(k)\omega_{ei}(k)\leq\bar{\omega}_{3}$, $\delta_{i}(k)=0$. Meanwhile, the volume of the ellipsoid $r_{i}(k)^{\top}\Pi r_{i}(k) = 1$ is proportional to $(\det[\Pi])^{-1/2}$, and  $(\det[\Pi])^{-1/2}$ shares the same minimizer with $-\log\det[\Pi]$. Therefore, the objective function in \eqref{lmipi} is chosen as $-\log\det[\Pi]$ as it is convex in $\Pi$ and leads to the tightest ellipsoidal bound (in terms of volume).
\end{remark}
\begin{remark}
Monitor \eqref{monitor} can be used by CAVs to detect system faults (sornsor, actuator, and dynamics faults) and \emph{non-stealthy} false data injection attacks, which covers a large class of anomalous behaviors. However, by definition, no monitor (including \eqref{monitor}) can identify stealthy FDI attacks. This is schematically depicted in Figure \ref{fig:x}.
\end{remark}

	\section{Attacker's Reachable Set: Security Metric 1}\label{reachable}
In this section, we aim to quantify the impact of stealthy attacks on the vehicle state of the vehicle $i$ when monitor \eqref{monitor} is used to detect attacks. We assume the attacker has knowledge of the vehicle model and the detector structure so that the injected $\delta_{i}$ does not trigger alarms by monitor \eqref{monitor}. This class of stealthy attacks can be characterized as a constrained control problem in $\delta_{i}$:
	\begin{equation}\label{stealthy}
		\left\lbrace \delta_{i} \in \mathbb{R} \hspace{1mm} \vline \hspace{1mm} \text{$r_{i}(k)$ satisfies \eqref{er} and }r_{i}(k)^{\top}\Pi r_{i}(k)\leq 1\right\rbrace.
	\end{equation}
	Because $\bar{L}B_{e1}=B_{e1}$ by construction, the closed-loop systems \eqref{ssd}, \eqref{sseee} can be written as follows:
	\begin{equation}\label{closed}
		\left\{\begin{split}
		&x_{i}(k+1)=Ax_{i}(k)+B\tilde{\omega}_{i}(k) + \Gamma\delta_{i}(k),\\
		&e_{ei}(k+1)=(I-LC_{e})A_{e}e_{ei}(k)\\
        &\hspace{15mm}-B_{e1}\omega_{ui}(k)-B_{e1}\delta_{i}(k) - L\omega_{ei}(k+1).
		\end{split}
		\right.
	\end{equation}
We are interested in the state trajectories that the attacker can induce in the system by injecting attack signals satisfying \eqref{stealthy} into the communication channels. This motivates the following definition.
	\begin{definition}[Stealthy Reachable Set].
		The stealthy reachable set of vehicle $i$ at time-instant $k$, $\mathcal{R}_{k}^{x_{i}}$, is the set of all states reachable by system \eqref{closed} through all possible initial conditions and disturbances satisfying $\tilde{\omega}_{i}^{\top}\tilde{\omega}_{i}\leq\bar{\omega}_{1}$, $\omega_{ui}^{\top}\omega_{ui}\leq\bar{\omega}_{2}$, and $\omega_{ei}^{\top}\omega_{ei}\leq\bar{\omega}_{3}$ and attacks satisfying \eqref{stealthy}-\eqref{closed}, i.e.,
		\begin{equation}
\mathcal{R}_{k}^{x_{i}} := \left\{ x_{i} \in \mathbb{R}^{4} \left|
			\begin{split}
				 & x_{i} \text{ and } e_{ei} \hspace{1mm}\text{satisfy \eqref{closed}},\\ & \delta_{i}\hspace{1mm}\text{satisfies}\hspace{1mm}\eqref{stealthy},
				\tilde{\omega}_{i}^{\top}(k)\tilde{\omega}_{i}(k)\leq\bar{\omega}_{1},\\ &\omega_{ui}(k)^{\top}\omega_{ui}(k)\leq\bar{\omega}_{2},\\ &\omega_{ei}(k)^{\top}\omega_{ei}(k)\leq\bar{\omega}_{3}.
			\end{split}
\right. \right\}.
		\end{equation}
	\end{definition}
	
Similar to \cite{murguia2020security}, we use the volume of the set $\mathcal{R}_{k}^{x_{i}}$ as the first security metric to assess the resiliency of CAVs to stealthy attacks on V2V communication networks. Since it is in general not tractable to compute $\mathcal{R}_{k}^{x_{i}}$ exactly, we look for an outer ellipsoidal approximation of the form $\mathcal{E}_{k}^{x_{i}} := \left\lbrace x_{i}\in\mathbb{R}^{4}|\hspace{1mm}x_{i}^{\top}P^{x_{i}}x_{i}\leq\alpha_{k}^{x_{i}}\right\rbrace $ such that $\mathcal{R}_{k}^{x_{i}}\subseteq\mathcal{E}_{k}^{x_{i}}$ for all $k \in \mathbb{N}$. This means that the ellipsoid $x_{i}^{\top}P^{x}x_{i}=\alpha_{k}^{x_{i}}$ contains all the possible trajectories induced by stealthy attacks satisfying \eqref{stealthy}. Since $\mathcal{E}_{k}^{x_{i}}$ is a good approximation of $\mathcal{R}_{k}^{x_{i}}$ for LTI systems and it can be computed efficiently using LMIs, we use the volume of $\mathcal{E}_{k}^{x_{i}}$ to approximate the security metric (the volume of $\mathcal{R}_{k}^{x_{i}}$).

To be able to use the results introduced in Lemma \ref{lemma1a} to compute these ellipsoidal approximations, we have to reformulate \eqref{closed} in the form of \eqref{s1}. In what follows, we present such a reformulation. Consider $r_{i}(k+1)$ in \eqref{er}, and compute $r_{i}(k+2)$:
		\begin{equation}\label{r}
		\begin{split}
			r_{i}(k+2)=&C_{e}A_{e}e_{ei}(k+1)+ \omega_{ei}(k+2),\\
			=&C_{e}A_{e}\bar{A}e_{ei}(k)-C_{e}A_{e}B_{e1}\omega_{ui}(k) +\omega_{ei}(k+2)\\ &- C_{e}A_{e}L\omega_{ei}(k+1)
			- C_{e}A_{e}B_{e1}\delta_{i}(k).
		\end{split}
	\end{equation}
Since $C_{e}A_{e}B_{e1}$ is full column rank by construction, we can compute $\delta_{i}(k)$ from \eqref{r} as:
	\begin{equation}\label{delta}
		\begin{split}
			\delta_{i}(k)=&-(C_{e}A_{e}B_{e1})^{\dagger}\big(r_{i}(k+2)-C_{e}A_{e}\bar{A}e_{ei}(k)\\
			 -&\omega_{ei}(k+2)+C_{e}A_{e}B_{e1}\omega_{ui}(k)
			+C_{e}A_{e}L\omega_{ei}(k+1) \big),
		\end{split}
	\end{equation}
	where $(C_{e}A_{e}B_{e1})^{\dagger}$ denotes the Moore-Penrose inverse of $C_{e}A_{e}B_{e1}$. Substituting \eqref{delta} into \eqref{closed} yields
	\begin{equation}\label{xc}
		\begin{split}
			&x_{i}(k+1)=Ax_{i}(k)+\Gamma(C_{e}A_{e}B_{e1})^{\dagger}C_{e}A_{e}\bar{A}e_{ei}(k)\\
			&+B\tilde{\omega}_{i}(k)-\Gamma\omega_{ui}(k)-\Gamma(C_{e}A_{e}B_{e1})^{\dagger}C_{e}A_{e}L\omega_{ei}(k+1)\\
&+\Gamma(C_{e}A_{e}B_{e1})^{\dagger}\omega_{ei}(k+2)-\Gamma(C_{e}A_{e}B_{e1})^{\dagger}r_{i}(k+2),
		\end{split}
	\end{equation}
	\begin{equation}\label{ec}
		\begin{split}
			&e_{ei}(k+1)=\big(I-B_{e1}(C_{e}A_{e}B_{e1})^{\dagger}C_{e}A_{e}\big)\bar{A}e_{ei}(k)\\
			&+\big(B_{e1}(C_{e}A_{e}B_{e1})^{\dagger}C_{e}A_{e}-I\big)L\omega_{ei}(k+1)\\
			&-B_{e1}(C_{e}A_{e}B_{e1})^{\dagger}\omega_{ei}(k+2)+B_{e1}(C_{e}A_{e}B_{e1})^{\dagger}r_{i}(k+2).
		\end{split}
	\end{equation}
	Define the extended state $\zeta_{i} :=\begin{bmatrix}
		x_{i}^{\top}&e_{ei}^{\top}
	\end{bmatrix}^{\top}$, the closed-loop dynamics \eqref{closed} can be written compactly as follows:
	\begin{equation}\label{zeta}
		\begin{split}
					\zeta_{i}(k+1)=&\mathcal{A}\zeta_{i}(k)+\mathcal{B}_{1}\tilde{\omega}_{i}(k)+\mathcal{B}_{2}\omega_{ui}(k)+\mathcal{B}_{3}\omega_{ei}(k+1)\\
			&+\mathcal{B}_{4}\omega_{ei}(k+2)+\mathcal{B}_{5}r_{i}(k+2),
		\end{split}
	\end{equation}
	with
	\begin{equation}\label{matrices}
		\begin{split}
			\mathcal{A}=&\begin{bmatrix}
				A&\Gamma(C_{e}A_{e}B_{e1})^{\dagger}C_{e}A_{e}\bar{A}\\
				\mathbf{0}&(I-B_{e1}(C_{e}A_{e}B_{e1})^{\dagger}C_{e}A_{e})\bar{A}
			\end{bmatrix},\\
			\mathcal{B}_{1}=&\begin{bmatrix}
				B\\
				\mathbf{0}
			\end{bmatrix}, \mathcal{B}_{2}=\begin{bmatrix}
				-\Gamma\\
				\mathbf{0}
			\end{bmatrix},\\
			\mathcal{B}_{3}=&\begin{bmatrix}
				-\Gamma(C_{e}A_{e}B_{e1})^{\dagger}C_{e}A_{e}L\\
				\big(B_{e1}(C_{e}A_{e}B_{e1})^{\dagger}C_{e}A_{e}-I\big)L
			\end{bmatrix},\\
			\mathcal{B}_{4}=&\begin{bmatrix}
			\Gamma(C_{e}A_{e}B_{e1})^{\dagger}\\
			-B_{e1}(C_{e}A_{e}B_{e1})^{\dagger}
			\end{bmatrix}, \mathcal{B}_{5}=\begin{bmatrix}
				-\Gamma(C_{e}A_{e}B_{e1})^{\dagger}\\
				B_{e1}(C_{e}A_{e}B_{e1})^{\dagger}
			\end{bmatrix}.
		\end{split}
	\end{equation}
The closed-loop dynamics \eqref{zeta} is now written in the form \eqref{s1} driven by peak-bounded perturbations $\tilde{\omega}_{i}$, $\omega_{ui}$, $\omega_{ei}$, and $r_{i}$. The residual $r_{i}$ acts as a peak-bounded perturbation because the attack sequence $\delta_i$ aims to be stealthy to the system monitor, which implies that $\delta_i$ satisfies \eqref{stealthy} (i.e., $r_i(k)^\top \Pi r_i(k) \leq 1$ for all $k \in \mathbb{N}$).

Define the reachable set for \eqref{zeta}:
\begin{equation}\label{rzeta}
\mathcal{R}_{k}^{\zeta_{i}} := \left\{ \zeta_{i} \in  \mathbb{R}^{10} \left|
			\begin{split}
				 & \zeta_{i}\hspace{1mm}\text{satisfies \eqref{zeta}}, r_i(k)^\top \Pi r_i(k) \leq 1,\\
				 &\tilde{\omega}_{i}(k)^{\top}\tilde{\omega}_{i}(k)\leq\bar{\omega}_{1},\\ &\omega_{ui}(k)^{\top}\omega_{ui}(k) \leq \bar{\omega}_{2},\\ &\omega_{ei}(k)^{\top}\omega_{ei}(k) \leq \bar{\omega}_{3}, k \in \mathbb{N}.
			\end{split}
\right. \right\}.
\end{equation}
We first use Corollary 1 to obtain outer approximations of the form $\mathcal{E}_{k}^{\zeta_{i}}=\left\lbrace \zeta_{i}\in\mathbb{R}^{10}|\zeta_{i}^{\top}P^{\zeta}\zeta_{i}\leq\alpha_{k}^{\zeta_{i}}\right\rbrace $ such that $\mathcal{R}_{k}^{\zeta_{i}}\subseteq\mathcal{E}_{k}^{\zeta_{i}}$. Then we project $\mathcal{E}_{k}^{\zeta_{i}}$ onto the $x_{i}$-hyperplane to obtain $\mathcal{E}_{k}^{x_{i}}$ satisfying $\mathcal{R}_{k}^{x_{i}}\subseteq\mathcal{E}_{k}^{x_{i}}$.
	\begin{theorem}
		Consider the closed-loop dynamics \eqref{zeta}-\eqref{matrices} with system matrices $(A,B,C,A_{e},B_{e1},C_{e})$, estimator gain $L$, controller gain $K$, monitor matrix $\Pi$, and perturbation bounds $\bar{\omega}_{1},\bar{\omega}_{2},\bar{\omega}_{3}\in\mathbb{R}_{>0}$. For a given $a\in(0,1)$, if there exist constants $a_{1}=a_{1}^{*}$, $a_{2}=a_{2}^{*}$, $a_{3}=a_{3}^{*}$, $a_{4}=a_{4}^{*}$, $a_{5}=a_{5}^{*}$, and matrix $P=P^{*}$ solution of
		\begin{equation}\label{lmi}
			\left\{\begin{split}
				&\min_{P,a_{1},\ldots, a_{5}}-\log\det[P],\\
				&s.t. \hspace{1mm}a_{1},a_{2}, a_{3}, a_{4}, a_{5}\in(0,1), a_{1}+a_{2}+a_{3}+ a_{4}+a_{5}\geq a,\\
				&P>0, \begin{bmatrix}
					aP&\mathcal{A}^{\top}P&\mathbf{0}\\
					P\mathcal{A}&P&P\mathcal{B}\\
					\mathbf{0}&\mathcal{B}^{\top}P&W_{a}
				\end{bmatrix}\geq 0,
			\end{split}\right.
		\end{equation}
		with $W_{a}:=\diag\begin{bmatrix}
			(1-a_{1})W_{1},\ldots,(1-a_{5})W_{5}
		\end{bmatrix}\in\mathbb{R}^{19\times 19}$, $\mathcal{B}:=(\mathcal{B}_{1},\mathcal{B}_{2},\mathcal{B}_{3},\mathcal{B}_{4},\mathcal{B}_{5})\in\mathbb{R}^{10\times 19}, W_{1}=\frac{1}{\bar{w}_{1}}I_{3}, W_{2}=\frac{1}{\bar{w}_{2}}, W_{3}=\frac{1}{\bar{w}_{3}}I_{5},W_{4}=\frac{1}{\bar{w}_{3}}I_{5}, W_{5}=\Pi$; then for all $k \in \mathbb{N}$, $\mathcal{R}_{k}^{\zeta_{i}}\subseteq\mathcal{E}_{k}^{\zeta_{i}}:=\left\lbrace \zeta_{i}^{\top}P^{\zeta}\zeta_{i}\leq\alpha_{k}^{\zeta_{i}}\right\rbrace $, with $P^{\zeta}:=P^{*}$, and $\alpha_{k}^{\zeta_{i}}:=a^{k-1}\zeta_{i}(1)^{\top}P^{*}\zeta_{i}(1)+((5-a)(1-a^{k-1}))/(1-a)$, and ellipsoid $\mathcal{E}_{k}^{\zeta_{i}}$ has minimum volume in the sense of Corollary 1. \linebreak
	\end{theorem}
	\textit{Proof:} See Appendix \ref{th1}.
	\begin{corollary}
		Consider the matrix $P^{\zeta}$ and function $\alpha_{k}^{\zeta}$ obtained by solving \eqref{lmi}. Partition $P^{\zeta}$ as
		\begin{equation}
			P^{\zeta}=:\begin{bmatrix}
				P_{1}^{\zeta}&P_{2}^{\zeta}\\
				(P_{2}^{\zeta})^{\top}&P_{3}^{\zeta}
			\end{bmatrix},
		\end{equation}
		with $P_{1}^{\zeta}\in\mathbb{R}^{4\times 4}$, $P_{2}^{\zeta}\in\mathbb{R}^{4\times 6}$, and $P_{3}^{\zeta}\in\mathbb{R}^{6\times 6}$. Then, for $k \in \mathbb{N}$, $\mathcal{R}_{k}^{x_{i}}\subseteq \mathcal{E}_{k}^{x_{i}}:=\left\lbrace x_{i}\in\mathbb{R}^{4}|x_{i}^{\top}P^{x}x_{i}\leq\alpha_{k}^{x_{i}}\right\rbrace $ with $P^{x}:=P_{1}^{\zeta}-P_{2}^{\zeta}(P_{3}^{\zeta})^{-1}(P_{2}^{\zeta})^{\top}$ and $\alpha_{k}^{x_{i}}:=\alpha_{k}^{\zeta_{i}}$.
	\end{corollary}
	\textit{Proof:} See Appendix \ref{cor1}.
	\section{Distance to Critical States: Security Metric 2}\label{cri}
	In this section, we propose a second security metric to quantify vulnerability of CAVs against stealthy attacks on V2V communication networks. To this end, we introduce the notion of \emph{critical states} $\mathcal{C}^{x_{i}}$ -- states that, if reached, compromise the integrity or safe operation of the vehicles. Such a region may represent states in which, a collision between two neighboring vehicles occurs, or the velocity of the vehicle exceeds the maximum allowed value. So far, we have characterized all the states that attacks can induce in the platoon while being stealthy to the system monitor (the stealthy reachable set $\mathcal{R}_{k}^{x_{i}}$).

\begin{figure}[t]\centering
	\includegraphics[width=0.4\textwidth]{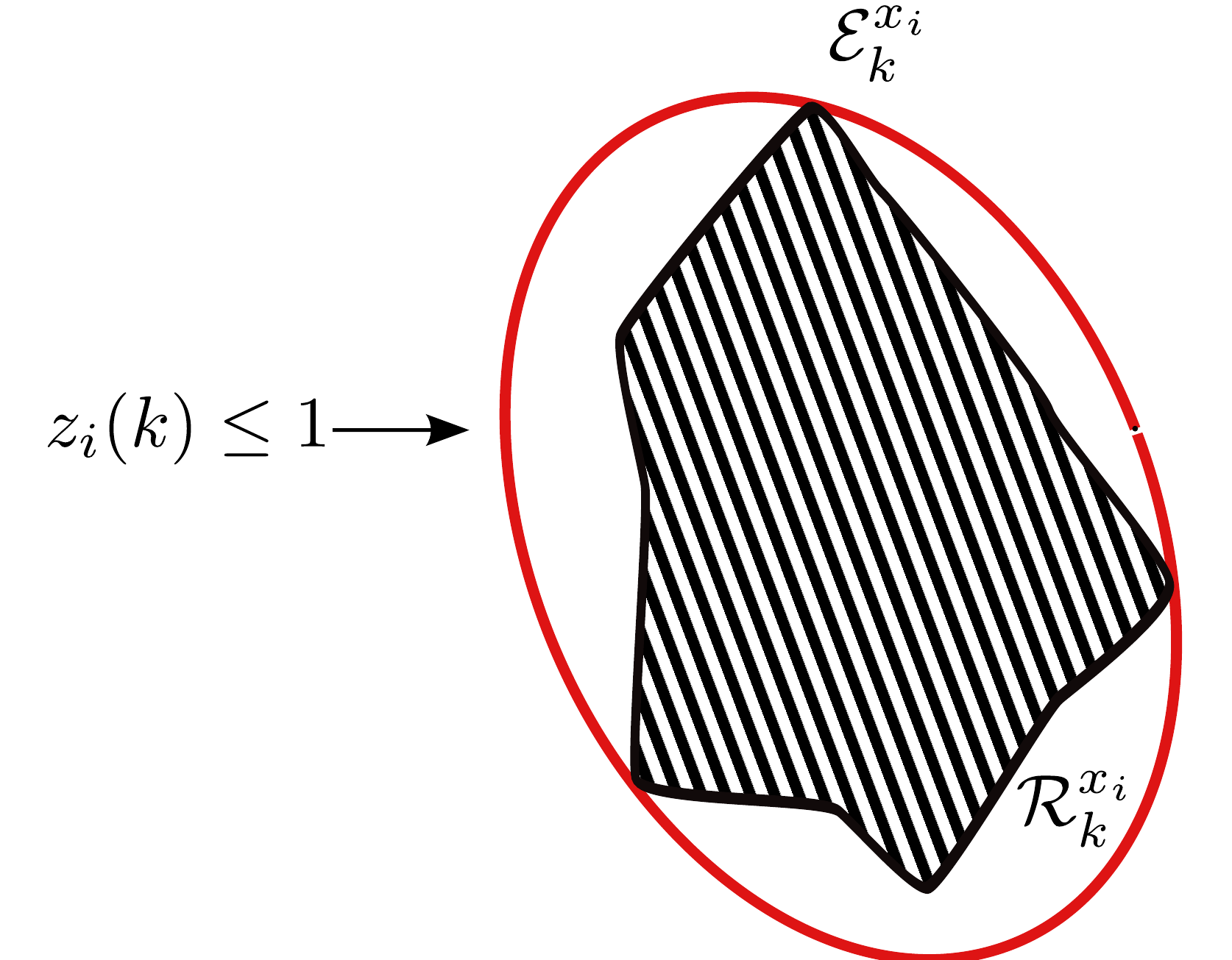}
	\caption{Stealthy reachable set of the $i$-th CAV $\mathcal{R}_{k}^{x_{i}}$ and outer ellipsoidal approximation $\mathcal{E}_{k}^{x_{i}}$ of $\mathcal{R}_{k}^{x_{i}}$.}
	\centering
	\label{fig:2}
\end{figure}

\begin{figure}[t]\centering
	\includegraphics[width=0.4\textwidth]{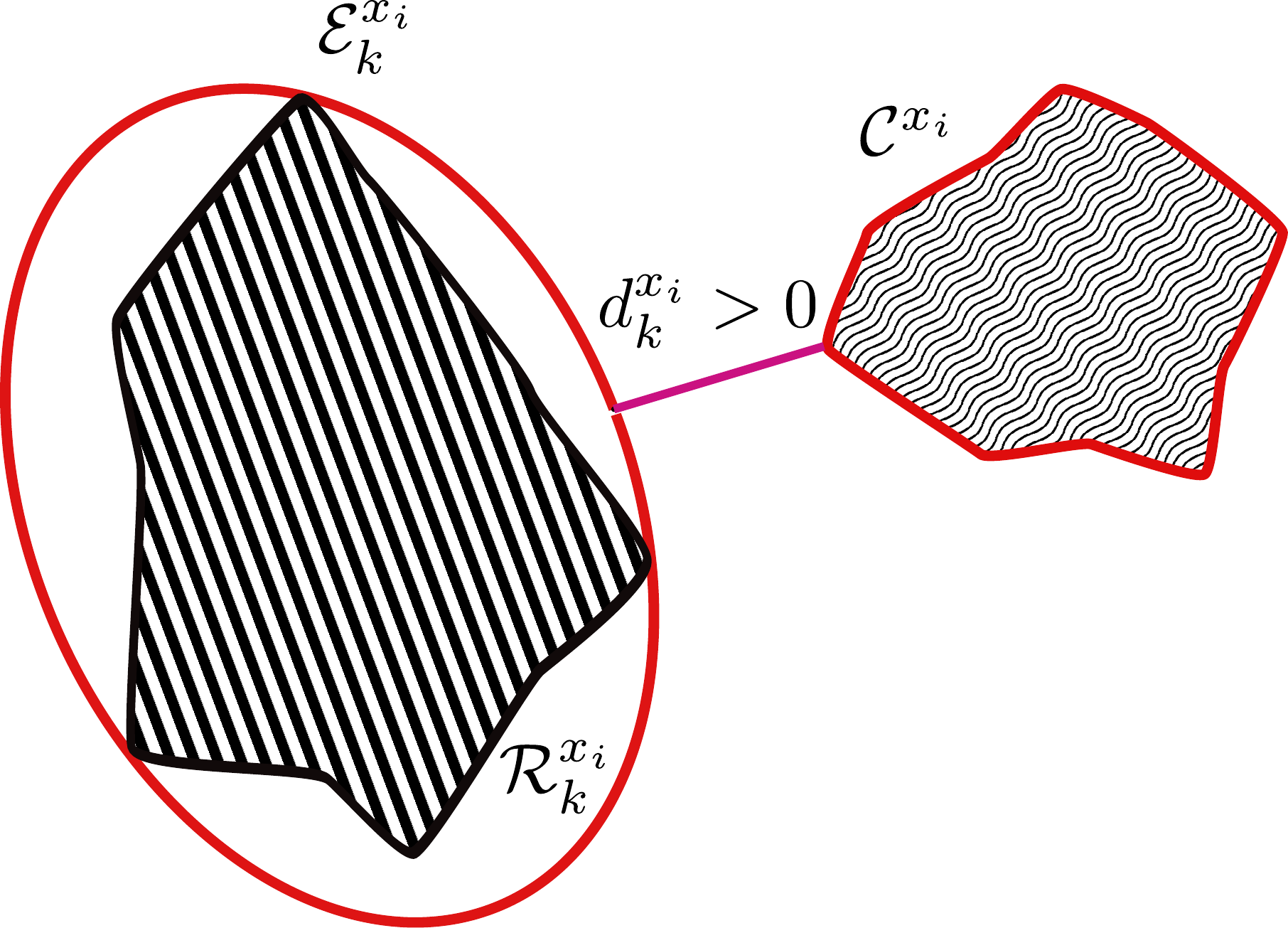}
	\caption{Minimum distance $d_{k}^{x_{i}}>0$ implies the $i$-th CAV is resilient to stealthy FDI attacks.}
	\centering
	\label{fig:3}
\end{figure}

\begin{figure}[t]\centering
	\includegraphics[width=0.4\textwidth]{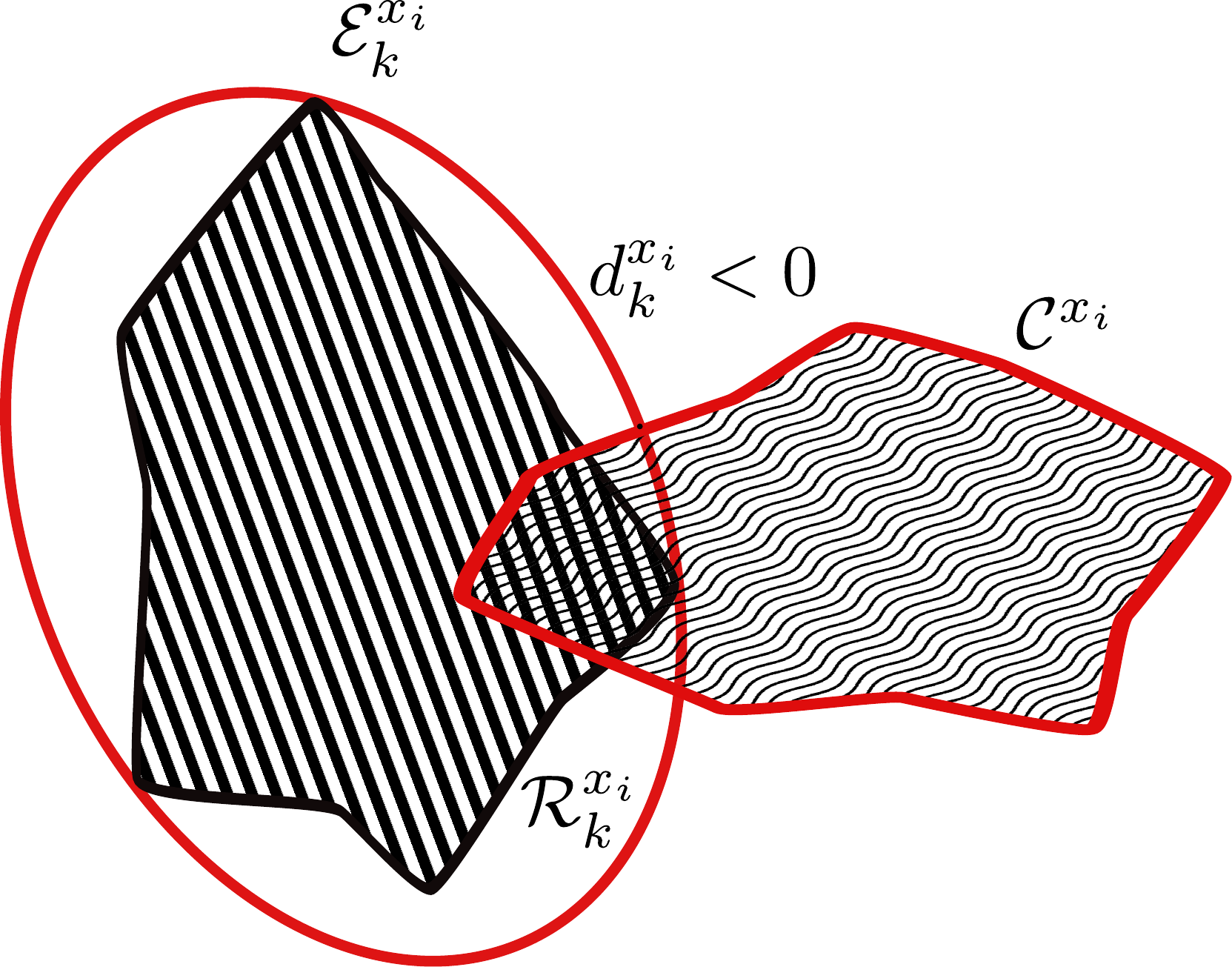}
	\caption{Minimum distance $d_{k}^{x_{i}}<0$ and there is an intersection between $\mathcal{R}_{k}^{x_{i}}$ and $\mathcal{C}^{x_{i}}$ -  the $i$-th CAV is vulnerable to stealthy FDI attacks.}
	\centering
	\label{fig:4}
\end{figure}

\begin{figure}[t]\centering
	\includegraphics[width=0.4\textwidth]{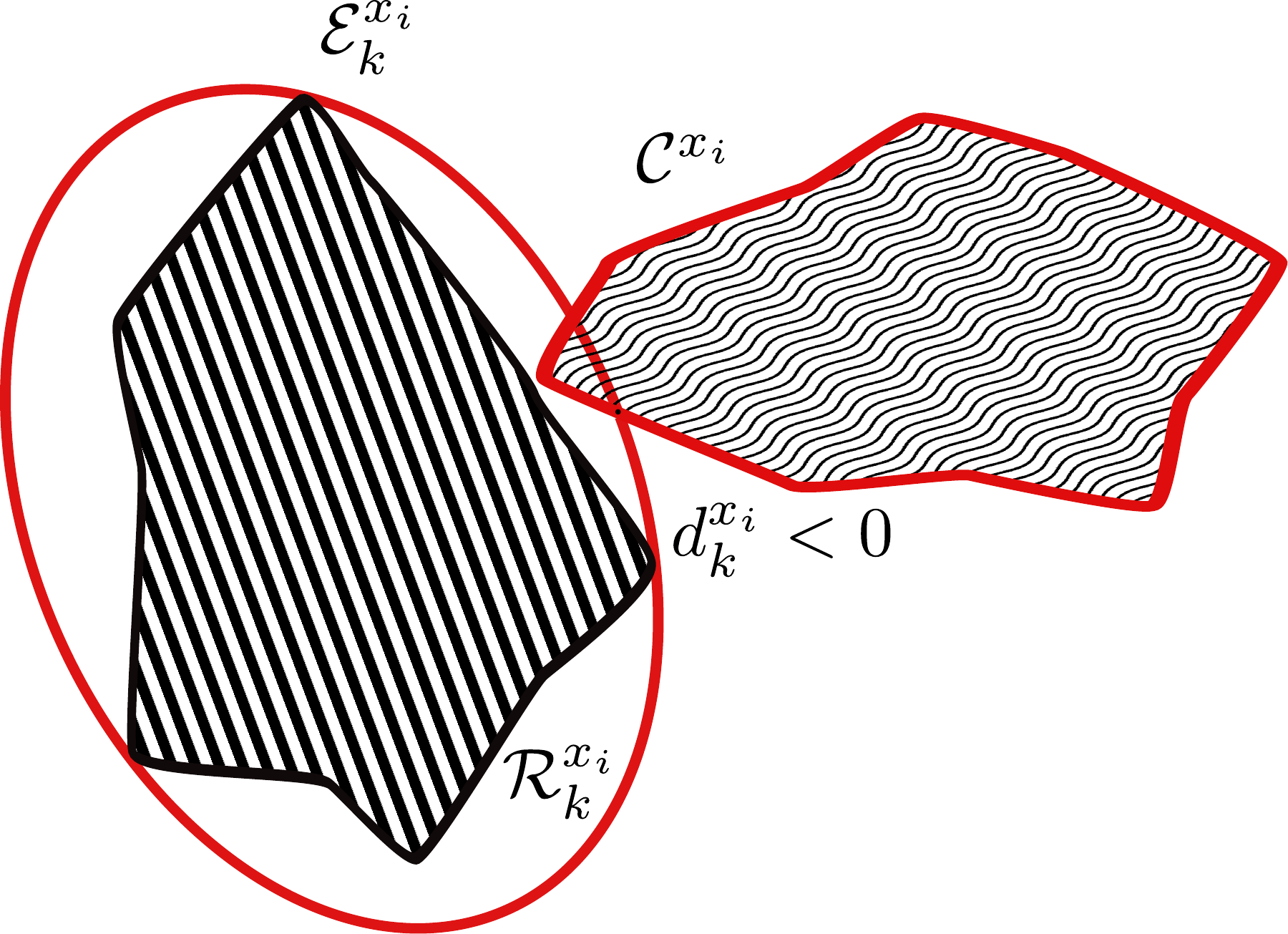}
	\caption{Minimum distance $d_{k}^{x_{i}}<0$ but there is no intersection between $\mathcal{R}_{k}^{x_{i}}$ and $\mathcal{C}^{x_{i}}$ -  the $i$-th CAV is resilient to stealthy FDI attacks.}
	\centering
	\label{fig:5}
\end{figure}

Therefore, if the intersection between $\mathcal{C}^{x_{i}}$ and $\mathcal{R}_{k}^{x_{i}}$ is not empty, there exist attack sequences $\delta_i$ that can induce a critical event by tampering with V2V communication networks. Conversely, if such an intersection is empty, we can ensure that attacks leading to critical events are impossible. Here, we compute the minimum distance $d_{k}^{x_{i}}$ from $\mathcal{E}_{k}^{x_{i}}$ to $\mathcal{C}^{x_{i}}$ and use it as an approximation of the distance between $\mathcal{R}_{k}^{x_{i}}$ and $\mathcal{C}^{x_{i}}$ (because computing $\mathcal{R}_{k}^{x_{i}}$ exactly is not tractable, we only have the approximation $\mathcal{E}_{k}^{x_{i}}$). The distance  $d_{k}^{x_{i}}$ gives us intuition of how far the actual reachable set  $\mathcal{R}_{k}^{x_{i}}$ is from $\mathcal{C}^{x_{i}}$.
From \eqref{e}, we have that the tracking error of each vehicle at the sampling time-instants is given by
	\begin{equation}
		e_{i}(k)=d_{i}(k)-(s_{i}+hv_{i}(k)).
	\end{equation}
	Since $d_{i}(k)< 0$ indicates a collision between vehicles $i$ and $i-1$, one subset of critical states is clearly given by:
	\begin{equation}\label{c1}
	\begin{split}
			e_{i}(k)<-s_{i}-hv_{i}(k).
		\end{split}
	\end{equation}
	Another potential subset of critical states is that the maximum allowed velocity $v_{\max}$ is exceeded, i.e.,
	\begin{equation}\label{c2}
		v_{i}(k) > v_{\max}.
	\end{equation}
	Combining \eqref{c1} and \eqref{c2}, the set of critical states is characterized by the union of half-spaces:
	\begin{equation}\label{c}
		\mathcal{C}^{x_{i}}:=\big\lbrace x_{i}\in\mathbb{R}^{4}\vline \hspace{1mm}c_{1}^{\top}x_{i}> s_{i} \text{ and } c_{2}^{\top}x_{i}> v_{\max}\big\rbrace,
	\end{equation}
	with $c_{1}^{\top}=\begin{bmatrix}
		-1&-h&0&0
	\end{bmatrix}$, and $c_{2}^{\top}=\begin{bmatrix}
		0&1&0&0
	\end{bmatrix}$.
	\begin{corollary}
		Consider the set of critical states $\mathcal{C}^{x_{i}}$ defined in \eqref{c} and the matrix $P^{x}$ and the function $\alpha_{k}^{x_{i}}$ obtained in Theorem 1. The minimum distance, $d_{k}^{x_{i}}$, between the outer ellipsoidal approximation of $\mathcal{R}_{k}^{x_{i}}$, $\mathcal{E}_{k}^{x_{i}}=\left\lbrace x_{i}\in\mathbb{R}^{4}\vline \hspace{1mm}x_{i}^{\top}P^{x}x_{i}\leq\alpha_{k}^{x_{i}}\right\rbrace $, and $\mathcal{C}^{x_{i}}$ is given by
		\begin{equation}\label{dd}
			d_{k}^{x_{i}}=\min\left(\frac{|b_{j}|-\sqrt{c_{j}^{\top}(P^{x})^{-1}c_{j}/\alpha_{k}^{x_{i}}}}{c_{j}^{\top}c_{j}} \right) , j=1,2.
		\end{equation}
	\end{corollary}
	\textit{Proof:} The minimum distance between an ellipsoid centered at the origin $\left\lbrace x\in\mathbb{R}^{n}|x^{\top}Px=1\right\rbrace $, $P>0$, and a hyperplane $\left\lbrace x\in\mathbb{R}^{n}|\hspace{1mm}c^{\top}x=b\right\rbrace $, $c\in\mathbb{R}^{n}$, $b\in\mathbb{R}$ is given by $(|b|-\sqrt{c^{\top}P^{-1}c})/c^{\top}c$ \cite{kurzhanskiy2006ellipsoidal}. The minimum distance between $\mathcal{C}^{x_{i}}$ and $\mathcal{E}^{x_{i}}_{k}$ is hence given by \eqref{dd}.\hfill$\blacksquare$
	\begin{remark}
		If $d_{k}^{x_{i}}> 0$ for all $k \in \mathbb{N}$, the ellipsoid $\mathcal{E}_{k}^{x_{i}}$ does not intersect with the set of critical states $\mathcal{C}^{x_{i}}$. Since $\mathcal{R}_{k}^{x_{i}}\subseteq\mathcal{E}_{k}^{x_{i}}$, the intersection between $\mathcal{C}^{x_{i}}$ and $\mathcal{R}_{k}^{x_{i}}$ is also empty. This indicates the $i$-th CAV will not collide with the preceding vehicle and its velocity will not exceed $v_{\max}$ under stealthy attacks. Hence, the $i$-th CAV is resilient to stealthy attacks on vehicular networks. A schematic representation of this idea is given in Figures \ref{fig:2}-\ref{fig:3}. On the other hand, if $d_{k}^{x_{i}}< 0$ for some $k \in \mathbb{N}$, there is a nonempty intersection between the ellipsoid $\mathcal{E}_{k}^{x_{i}}$ and the set of critical states $\mathcal{C}^{x_{i}}$. This indicates that the $i$-th CAV might collide with the preceding vehicle or exceed the maximum allowed velocity under stealthy attacks. The latter case is depicted in Figure \ref{fig:4}. However, due to the potential conservatism of the ellipsoidal bounds, the scenario depicted in Figure \ref{fig:5} may also occur, i.e., $d_{k}^{x_{i}}<0$ but there is an empty intersection between $\mathcal{R}_{k}^{x_{i}}$ and $\mathcal{C}^{x_{i}}$. Therefore, if $d_{k}^{x_{i}}> 0$, the $i$-th CAV is risk-free in the presence of stealthy FDI attacks; otherwise, the $i$-th CAV is under risk. Hence, a controller redesign might be necessary to make $d_{k}^{x_{i}}> 0$ for all $k \in \mathbb{N}$ so that potential security risks can be removed.
	\end{remark}

	\section{Simulation Experiments}\label{simulation}
	In this section, we conduct a simulation study to illustrate the effectiveness of our tools.
	
\begin{figure}[t]\centering
	\includegraphics[width=0.5\textwidth]{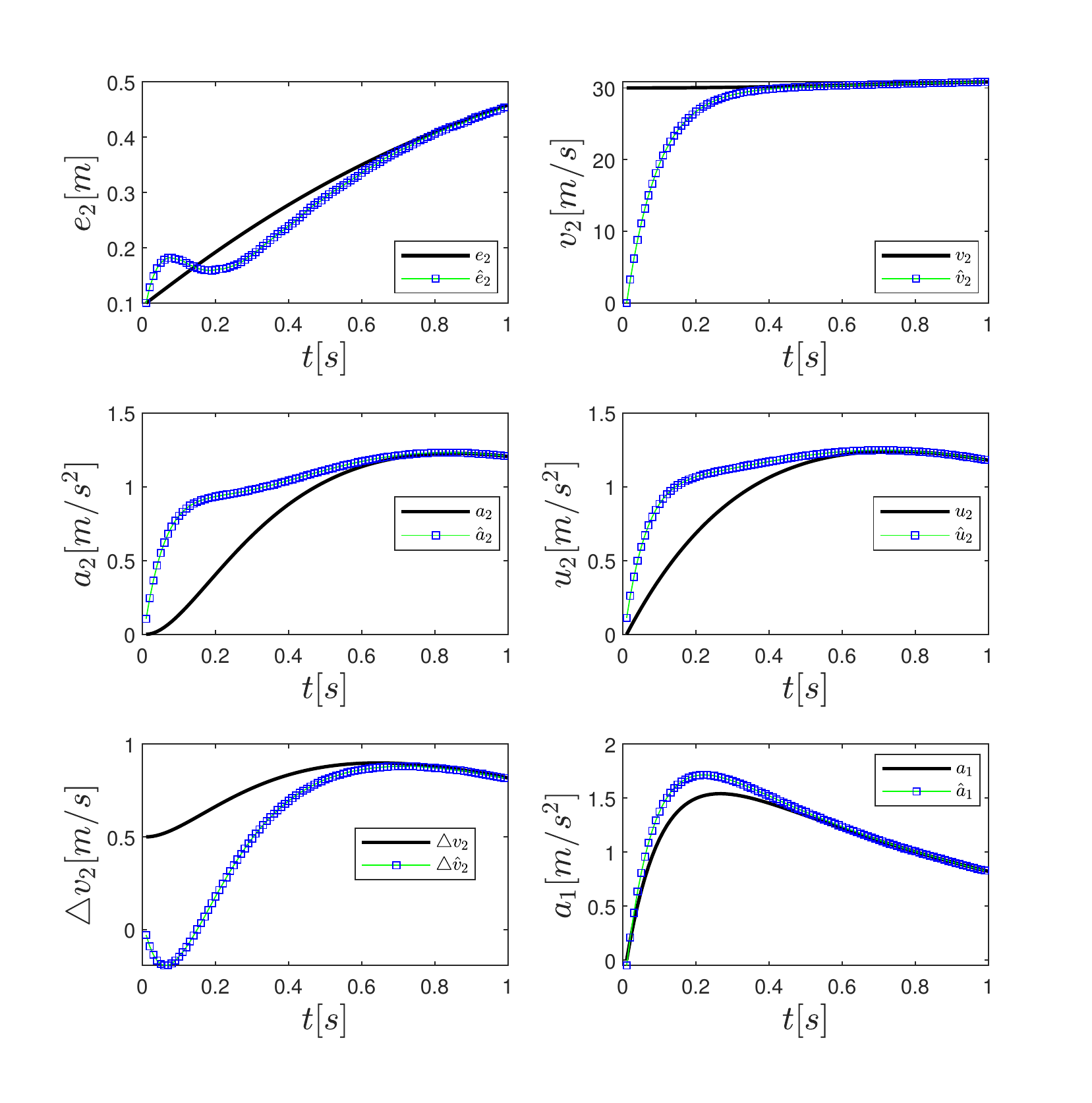}
	\caption{The estimate $\hat{x}_{e2}$ converges to the true vehicle state $x_{e2}$.}
	\centering
	\label{fig:e}
\end{figure}

		\textbf{Example 1.}
	Consider two homogeneous vehicles in a platoon. Suppose $h=0.5$, $\tau=0.1$, $K=\begin{bmatrix}
		0.2&0.7
	\end{bmatrix}$. We obtain the discrete-time system \eqref{ssd} with sampling interval $T_{s}=0.1$ seconds. At each time step $k$, the desired acceleration of vehicle 1, i.e., $u_{1}(k)$ is transmitted from vehicle $1$ to vehicle $2$ via the communication network. We next obtain the optimal state estimator using Lemma 3. By solving the LMIs in \eqref{lmi3} and performing a grid search over $\alpha\in(0,1)$, we obtain the optimal (smallest) ISS gain, $\gamma=1.0689$ with corresponding estimator matrix:
\begin{equation}\label{estimatorgain}
	\begin{split}
		&L=\\
		&\begin{bmatrix}
			0.1023&-0.0002&0.0082&0.0261&0.0057\\
			-0.0002&0.1126&0.0030&0.0031&-0.0000\\
			0.0082&0.0030&0.0429&0.0354&-0.0034\\
			0.0261&0.0031&0.0354&0.0331&-0.0021\\
			0.0057&-0.0000&-0.0034&-0.0021&0.1081\\
			-0.0031&-0.0017&0.0017&0.0003&0.0108
		\end{bmatrix}.
	\end{split}
\end{equation}
Let $u_{1}(k)=2e^{-0.01k}$, the initial relative velocity between vehicles 1 and 2 be $0.5m/s$, the initial spacing error be $0.1m$, and the initial velocity of vehicle 2 be $30m/s$. Therefore, $x_{e2}(1)=\begin{bmatrix}
	0.1&30&0&0&0.5&0
\end{bmatrix}^{\top}$. The initial condition $\hat{x}_{e2}(1)$ is randomly chosen. The performance of the estimator is shown in Figure \ref{fig:e}.

	\textbf{Example 2.}
	We assume each CAV in a platoon is equipped with the controller provided in \cite{ploeg2013lp}, an the standard fault detector that was introduced in Section \ref{estimation}. We use the approaches provided in Sections \ref{reachable} and \ref{cri} to approximate the security metrics. We show that our reachability analysis can be used to guide a controller redesign to reduce the impact of stealthy attacks on platooning performance.
	
\begin{figure}[t]\centering
	\includegraphics[width=0.4\textwidth]{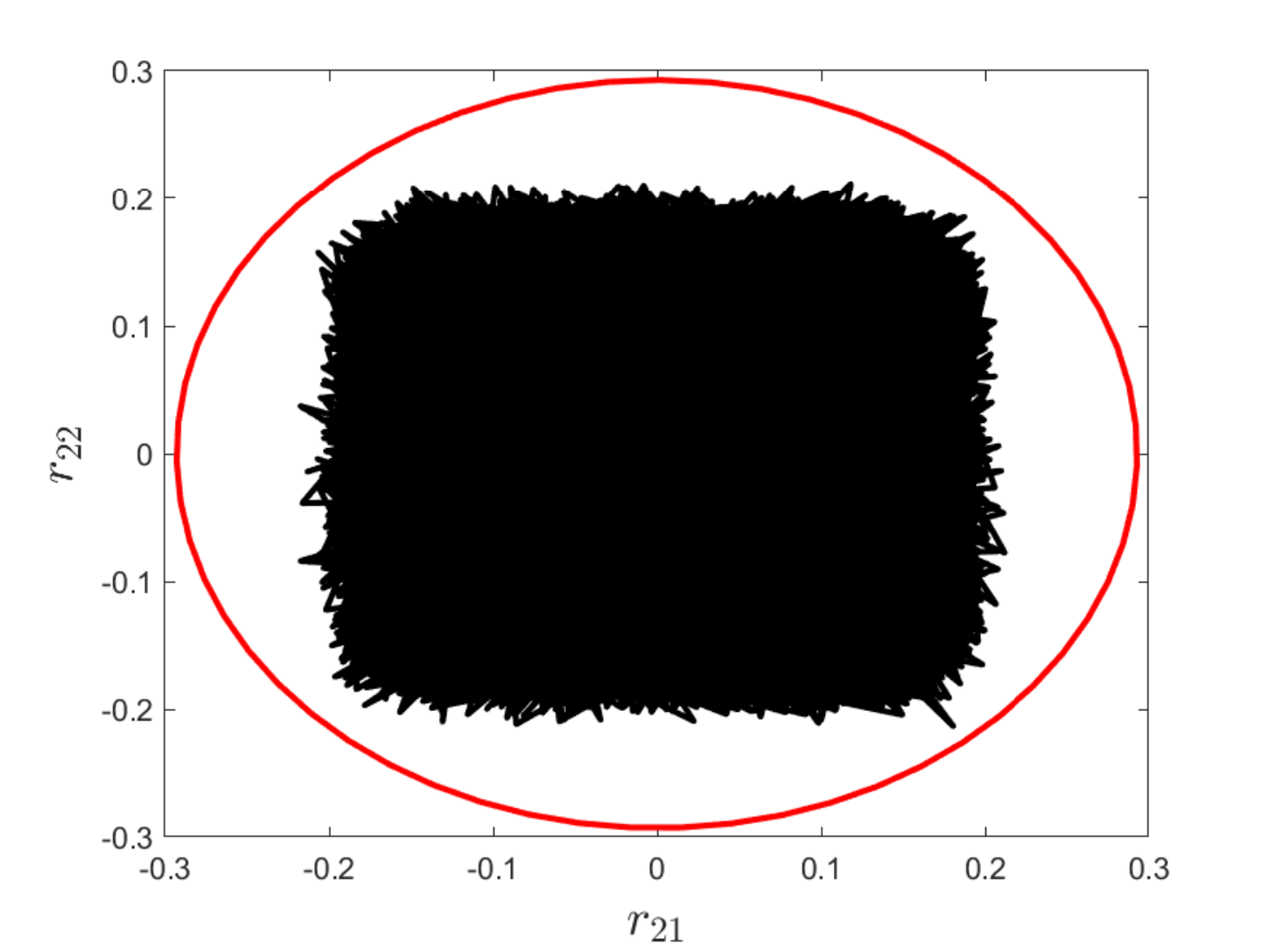}
	\caption{The projection of the ellipsoid $r_{2}(k)^{\top}\Pi r_{2}(k)=1$ onto the plane of the residual's first two elements $(r_{21},r_{22})$ (red) outer-bounds many (10,000) residual trajectories generated by a Monte-Carlo simulation (black).}
	\centering
	\label{fig:r}
\end{figure}

Consider two homogeneous vehicles in a platoon driving at a constant speed of $v=30m/s$ on the highway. We let $h=0.5$ and $\tau=0.1$, and obtain the discrete-time system \eqref{ssd} with sampling interval $T_{s}=0.1$ seconds. At each time step $k$, $u_{1}(k)$ is transmitted from vehicle $1$ to vehicle $2$ via the communication network, and the adversary injects $\delta_{1}(k)$ satisfying \eqref{stealthy} (stealthy attacks) to $u_{1}(k)$. We use the upper bounds on different vehicle sensors disturbances provided in \cite{de2017survey}. In particular, we take $||\omega_{d2}||_{\infty}= 0.1$, $||\omega_{v2}||_{\infty}= 0.01$,  $||\omega_{e2}||_{\infty}=0.1414$, and let $||\omega_{u2}||_{\infty}=0.01$. We assume the velocity of the vehicle $1$ can vary from $v_{\min}=0$ to $v_{\max}=35m/s$ and desired acceleration can vary from $u_{\min}=-2.5m/s^2$ to $u_{\max}=3m/s^2$, i.e.,  $v_{1} \in [0,35]$ and $u_{1}\in[-2.5, 3]$. Therefore, we have
\begin{equation*}
	\begin{split}
			\bar{\omega}_{1}=&||\omega_{d2}||_{\infty}^2+(v_{\max}+||\omega_{v2}||_{\infty})^{2}\\
			&+(\max\left\lbrace |u_{\min}|, |u_{\max}|\right\rbrace +||\omega_{u2}||_{\infty})^2=1234.8,\\
			\bar{\omega}_{2}=&||\omega_{u2}||_{\infty}^2=0.0001,\\
			\bar{\omega}_{3}=&||\omega_{e2}||_{\infty}^{2}=0.0200.
	\end{split}
\end{equation*}
We use the optimal estimator we obtain from Example 1 with $L$ given in \eqref{estimatorgain}.
We adopt the controller gain given in \cite{Ploeg2014} for fulfilling vehicle tracking objective and string stability with $K=\begin{bmatrix}
		0.2&0.7
	\end{bmatrix}$. Next, we obtain the monitor matrix $\Pi$ by solving the convex program in \eqref{lmipi} to obtain:
	\begin{equation}\label{pi}
	\begin{split}
		&\Pi=\\
		&\begin{bmatrix}
			11.6536&0.0002&0.0290&-0.1110&-0.0610\\
			0.0002&11.6527&-0.0580&-0.0000&0.0003\\
			0.0290&-0.0580&12.8425&-0.6275&0.0579\\
			-0.1110&-0.0000&-0.6275&11.9273&-0.0123\\
			-0.0610&0.0003&0.0579&-0.0123&11.6525
		\end{bmatrix}.
	\end{split}
\end{equation}
In order to illustrate the performance of the designed monitor visually, we project the monitor ellipsoid $r_{2}(k)^{\top}\Pi r_{2}(k)=1$ onto the plane of the residual's first two elements, $(r_{21},r_{22})$, and show a Monte-Carlo simulation of 10,000 trajectories of $\left\lbrace r_{21}(k)\right\rbrace ,\left\lbrace r_{22}(k)\right\rbrace $ generated by the residual dynamics \eqref{er}. With these simulations, we construct an empirical reachable set satisfying the corresponding bounds. The tightness of the ellipsoidal bound is shown in Figure \ref{fig:r}.

 Once we have obtained $L$ and $\Pi$ of the monitor and fixed $K$ from \cite{Ploeg2014}, we solve the optimization in \eqref{lmi} by performing a grid search over $a\in(0,1)$ and seeking for the $P^{*}=P^{\zeta_{2}}$ leading to the smallest $-\log(\det[P^{\zeta_{2}}])$. Then, we compute $P^{x}$ using the method provided in Corollary 1 and obtain $\mathcal{R}_{k}^{x_{2}}\subseteq\mathcal{E}_{k}^{x_{2}}:=\left\lbrace x_{2}\in\mathbb{R}^{4}|x_{2}^{\top}P^{x}x_{2}\leq\alpha_{k}^{x_{2}}\right\rbrace $ with
  		\begin{equation}
 	P^{x}=\begin{bmatrix}
 		0.0383&0.0189&-0.0413&-0.0007\\
 		0.0189&0.0104&-0.0233&0.0026\\
 		-0.0413&-0.0233&0.0776&-0.0313\\
 		-0.0007&0.0026&-0.0313&0.0321
 	\end{bmatrix}.
 \end{equation}
	Suppose $x_{2}(1)=\begin{bmatrix}
		0&30&0&0
	\end{bmatrix}^{\top}$. We let the critical states be characterized by \eqref{c}. We let $e_{e2}(1) = \mathbf{0}$, so no transients due to initial estimation errors. Thus, we
	have $\zeta_{2}(1)=\begin{bmatrix}
		0&30&0&0&0&0&0&0&0&0
	\end{bmatrix}^{\top}$. The average standstill distance of $s_{2}=3$ meters given in \cite{houchin2015measurement} is adopted. From \eqref{dd}, the minimum distance between $\mathcal{E}_{k}^{x_{2}}$ and $\mathcal{C}^{x_{2}}$ can be computed with $\alpha_{k}^{x_{2}}=a^{k-1}\zeta_{2}(1)^{\top}P^{*}\zeta_{2}(1)+((5-a)(1-a^{k-1}))/(1-a)$, $c_{1}^{\top}=\begin{bmatrix}
		-1&-h&0&0
	\end{bmatrix}$, and $c_{2}^{\top}=\begin{bmatrix}
		0&1&0&0
	\end{bmatrix}$, $b_{1}=s_{2}$, and $b_{2}=v_{\max}=35$. \\
We calculate the distance as
	\begin{equation}
		d_{jk}^{x_{2}}=\frac{|b_{j}|-\sqrt{c_{j}^{\top}(P^{x})^{-1}c_{j}/\alpha_{k}^{x_{2}}}}{c_{j}^{\top}c_{j}}, j\in\left\lbrace 1,2\right\rbrace,
	\end{equation}
for $k\geq 1$. We obtain that $d_{1k}^{x_2}$ and $d_{2k}^{x_2}$ are both positive for $k\geq 1$. Hence, $d_{k}^{x_{2}}$ as defined in \eqref{dd} is positive for $k\geq 0$.

Therefore, given standstill distance $s_{2}=3$ meters, the controller $K=\begin{bmatrix}
		0.2&0.7
	\end{bmatrix}$, estimator \eqref{sse} with $L$ given in \eqref{estimatorgain}, and monitor \eqref{monitor} with $\Pi$ given by \eqref{pi}, the intersection between the set of critical states and the stealthy reachable set is empty. This indicates that vehicle $2$ is not likely to collide with vehicle $1$ and velocity of vehicle $2$ will not exceed maximum allowed speed under stealthy attacks on the V2V communication networks. Hence, for this configuration, the CAVs are free of risk for critical stealthy attacks.

However, if we choose the controller gain $K$ not carefully enough, the CAVs might be under risk when the standstill distance of $s_{2}=3$ meters is adopted. For instance,
it has been verified in \cite{ploeg2011design}\cite{Ploeg2014} that any controller $K=\begin{bmatrix}
		k_{p}&k_{d}
	\end{bmatrix}$ satisfying $k_{p},k_{d}>0$ and $k_{d}>k_{p}\tau$ fulfills the tracking objective and string stability. If we choose $\bar{K}=\begin{bmatrix}
	0.9&0.1
\end{bmatrix}$, and obtain $L$ and $\Pi$ by solving the LMIs in \eqref{lmi3} and \eqref{lmipi} respectively, it can be verified that $d_{1k}^{x_{2}}< 0$ for all $k\leq 38$, i.e., there is an intersection between the set of critical states and the stealthy reachable set. This indicates that the CAVs are under risk for this configuration since a collision between vehicles $2$ and $1$ is feasible under stealthy attacks on the V2V communication networks.

\section{Conclusion}\label{conclusion}

For a standard CACC scheme for platooning, we have provided algorithmic tools for risk assessment of CAVs. First, we have given design methods to construct robust observer-based attack detection schemes in terms of the solution of a set of semidefinite programs. Our detectors provide optimal performance in the sense of minimizing the effect of noise and system disturbances in the detection procedure. However, because we consider a broad class of perturbations, intelligent attacks can surpass the detector by hiding within the system uncertainty. This class of attacks is what we refer to as stealthy attacks. Because stealthy attacks are constrained by the class of detector used, we characterize all the stealthy attacks that adversaries could launch in the system given the detector that we propose. We have proposed two security metrics to quantify how stealthy attacks on V2V communication networks can deteriorate the platooning performance. We have provided constructive methods (in terms of convex programs) to approximate these metrics given the system configuration. We showed how to use these tools for risk assessment given the system configuration. We remark that, as future work, our analysis approach can be used for guiding a controller redesign such that the potential impact of stealthy attacks can be minimized or avoided all together (as shown in the simulation section).

	\bibliographystyle{ieeetr}
	\bibliography{Observer1}
		\appendices
	\section{Proof of Lemma 3}\label{lm3}
Define $V(k)=e_{ei}(k)^{\top}Pe_{ei}(k)$. Evaluate the difference $\triangle V(k)\triangleq V(k+1)-V(k)$, we have
\begin{equation}
	\begin{split}
		&\triangle V=e_{ei}(k+1)^{\top}Pe_{ei}(k+1)-e_{ei}(k)^{\top}Pe_{ei}(k)\\
		&=e_{ei}(k)^{\top}\big(\bar{A}^{\top}P\bar{A}-P\big)e_{ei}(k)-2\omega_{ui}(k)^{\top}B_{e1}^{\top}P\bar{A}e_{ei}(k)\\
		&-2\omega_{ei}(k)^{\top}L^{\top}P\bar{A}e_{e1}(k)+2\omega_{ui}(k)^{\top}B_{e1}^{\top}PL\omega_{ei}(k+1)\\
		&+\omega_{ui}(k)^{\top}B_{e1}^{\top}PB_{e1}\omega_{ui}(k)+\omega_{ei}(k+1)^{\top}L^{\top}PL\omega_{ei}(k+1)
	\end{split}
\end{equation}
Since $P>0$, take Schur complements of \eqref{eq:l1} and we have
\begin{equation}\label{matrix1}
	\Xi+\Omega_{12}^{\top}P^{-1}\Omega_{12}\leq 0,
\end{equation}
where
\begin{equation}
	\Xi=\begin{bmatrix}
		(\alpha-1)P&*&*\\0&-\alpha I&*\\0&0&-\alpha I
	\end{bmatrix},
\end{equation}
and
\begin{equation}
	\Omega_{12}=\begin{bmatrix}
		(P-YC)A_{e}&PB&Y
	\end{bmatrix}.
\end{equation}
Compute \eqref{matrix1} and let $Y=PL$, we have
\begin{equation}\label{ox}
	\Omega+X\leq 0
\end{equation}
with
\begin{equation}
	\begin{split}
		\Omega=\begin{bmatrix}
			\bar{A}^{\top}P\bar{A}-P&*&*\\
			B_{e1}^{\top}P\bar{A}&B_{e1}^{\top}PB_{e1}&*\\
			L^{\top}P\bar{A}&L^{\top}PB_{e1}&L^{\top}PL
		\end{bmatrix}
	\end{split}
\end{equation}
and
\begin{equation}
	\begin{split}
		X=\begin{bmatrix}
			\alpha P&*&*\\
			0&-\alpha I&0\\
			0&*&-\alpha I
		\end{bmatrix}.
	\end{split}
\end{equation}
Taking a congrence transformation of \eqref{ox} with $\sigma_{i}(k)=\begin{bmatrix}
	e_{ei}(k)^{\top}&\omega_{ui}(k)^{\top}&\omega_{ei}(k+1)^{\top}
\end{bmatrix}^{\top}$ yields
\begin{equation}
	\begin{split}
			&\triangle V(k)+\alpha V(k)\\
			&-\alpha\mu_{1}\big(\omega_{ui}(k)^{\top}\omega_{ui}(k)+\omega_{ei}(k+1)^{\top}\omega_{ei}(k+1)\big)\leq 0,
	\end{split}
\end{equation}
for some $\alpha\in(0,1)$, and all $k\geq 0$, from which we can obtain
\begin{equation}
	V(k)\leq(1-\alpha)^{k}V(0)+\mu_{1}(||\omega_{ui}||_{\infty}^{2}+||\omega_{ei}||_{\infty}^{2}).
\end{equation}
Then it is satisfied that
\begin{equation}\label{x1}
	\sqrt{V(k)}\leq\sqrt{(1-\alpha)^{k}\lambda_{\max}}|e_{ei}(0)|+\sqrt{\mu_{1}}(||\omega_{ui}||_{\infty}+||\omega_{ei}||_{\infty}),
\end{equation}
where $\lambda_{\max}$ denotes the largest eigenvalue of $P$.
If \eqref{eq:l2} is satisfied, then
\begin{equation}\label{x2}
	|e_{ei}(k)|\leq\sqrt{\mu_{2}V(k)}.
\end{equation}
From \eqref{x1} and \eqref{x2}, we have
\begin{equation}\label{iss2}
	\begin{split}		|e_{ei}(k)|\leq&\sqrt{\mu_{2}\lambda_{\max}(1-\alpha)^{k}}|e_{ei}(0)|\\
		&+\sqrt{\mu_{1}\mu_{2}}(||\omega_{ui}||_{\infty}+||\omega_{ei}||_{\infty}). k\geq 0.
	\end{split}
\end{equation}
\eqref{iss2} is of the form \eqref{iss} with $c=\sqrt{\mu_{2}\lambda_{\max}}$ and $\lambda=\sqrt{1-\alpha}$. Moreover, by minimizing $\mu_{1}+\mu_{2}$, we can minimize the upper bound on $\gamma$ since $\gamma=\sqrt{\mu_{1}\mu_{2}}\leq \frac{1}{2}(\mu_{1}+\mu_{2})$. Hence, the effect of disturbances on the estimation error is minimized. \hfill$\blacksquare$
\section{Proof of Proposition 1}\label{p1}
From Lemma 3 (see \eqref{iss}), in steady state, the estimation error satisfies
\begin{equation}
	|e_{ei}(k)|^{2}\leq\gamma^{2}(\bar{\omega}_{2}+\bar{\omega}_{3}).
\end{equation}
By \eqref{er}, the monitor inequality, $r_{i}(k+1)^{\top}\Pi r_{i}(k+1) = 1$, can be written, in steady state, in terms of the estimation error and the perturbations as follows
\[
\begin{split}
	&(C_{e}A_{e}e_{ei}(k) +\omega_{ei}(k+1))^{\top} \Pi \\
	&\times (C_{e}A_{e}e_{ei}(k) +\omega_{ei}(k+1)) \leq 1.
\end{split}
\]
Then, by the $S$-procedure \cite{boyd1994linear}, if there exist $\lambda_{1}$, $\lambda_{2}\in\mathbb{R}_{>0}$ satisfying
\begin{equation}\label{rr}
	\begin{split}
		&\big(C_{e}A_{e}e_{ei}(k) +\omega_{ei}(k+1)\big)^{\top} \Pi \\
		&\times \big(C_{e}A_{e}e_{ei}(k) +\omega_{ei}(k+1)\big) - 1\\
		&-\lambda_{1}\big(e_{ei}(k)^{\top}e_{ei}(k)- \gamma^{2}(\bar{\omega}_{2}+\bar{\omega}_{3})\big)\\
		&-\lambda_{2}\big(\omega_{ei}(k+1)^{\top}\omega_{ei}(k+1)-\bar{\omega}_{3}\big)\leq 0;
	\end{split}
\end{equation}
then, $r_{i}(k+1)^{\top}\Pi r_{i}(k+1) \leq 1$ for all $e_{ei}(k)$, $\omega_{ei}(k)$, and $\omega_{ui}(k)$ satisfying $e_{ei}(k)^{\top}e_{ei}(k)\leq \gamma^{2}(\bar{\omega}_{2}+\bar{\omega}_{3})$, $\omega_{ei}(k)^\top\omega_{ei}(k) \leq \bar{\omega}_{3}$, and $\omega_{ui}(k)^{\top}\omega_{ui}(k)\leq\bar{\omega}_{2}$.
Define $\nu_{i}(k):=\begin{bmatrix}
	e_{ei}(k)^{\top}&\omega_{ei}(k+1)&1
\end{bmatrix}^{\top}$; then, \eqref{rr} can be written as
\begin{equation}\label{ine}
	\nu_{i}^{\top}\underbrace{\begin{bmatrix}
			f_{1}&*&*\\
			\Pi C_{e}A_{e}&f_{2}&*\\
			0&0&f_{3}
	\end{bmatrix}}_{Q}\nu_{i}\geq 0.
\end{equation}
with
\begin{equation}
	\begin{split}
		&f_{1}=\lambda_{1}I-A_{e}^{\top}C_{e}^{\top}\Pi C_{e}A_{e},\\
		&f_{2}=\lambda_{2}I-\Pi,\\
		&f_{3}=1-\lambda_{1}\gamma^{2}(\bar{\omega}_{2}+\bar{\omega}_{3})-\lambda_{2}\bar{\omega}_{3}.
	\end{split}
\end{equation}
Inequality \eqref{ine} is satisfied if and only if matrix $Q$ is positive semidefinite. Therefore, we have $r_{i}(k+1)^{\top}\Pi r_{i}(k+1)\leq 1$ in steady state if $Q \geq 0$. We minimize $\log\det[\Pi]^{-1}$ to make the ellipsoidal bound as tight as possible.\hfill$\blacksquare$
\section{Proof of Theorem 1}\label{th1}
The reachable set \eqref{rzeta} is the reachable set of system \eqref{zeta}, which is a LTI system driven by multiple peak-bounded perturbations. If \eqref{lmi} is satisfied, then, by Lemma \ref{lemma1a}, $\mathcal{E}_{k}^{\zeta_{i}}$ above contains the reachable set $\mathcal{R}_{k}^{\zeta_{i}}$ and this ellipsoid has minimum volume.\hfill$\blacksquare$
\section{Proof of Corollary 1}\label{cor1}
From Theorem 1, we have $\zeta_{i}^{\top}P^{\zeta}\zeta_{i}\leq\alpha_{k}^{\zeta_{i}}$ for $k \in \mathbb{N}$. From Lemma \ref{lemma1}, the projection of $\zeta_{i}^{\top}P^{\zeta}\zeta_{i}\leq\alpha_{k}^{\zeta_{i}}$ onto $x_{i}$-hyperplane is given by $\mathcal{E}_{k}^{x_{i}}$ above, and because $\mathcal{R}_{k}^{\zeta_{i}} \subseteq \mathcal{E}_{k}^{\zeta_{i}}$, then $\mathcal{R}_{k}^{x_{i}}\subseteq \mathcal{E}_{k}^{x_{i}}$. \hfill$\blacksquare$
\end{document}